%% file: main-document.tex
\documentclass{template/rsproca_new}

\titlehead{Research}

\usepackage{xcolor}
\usepackage{tikz}
\usepackage{cite}
\usepackage{pgfplots}
\usepackage{natbib}

\def\dimn#1{#1^*}                                       
\def\vect#1{\mathbf{#1}}                                
\def\eq#1{\begin{equation*} #1 \end{equation*}}         
\def\eqn#1{\begin{equation} #1 \end{equation}}          
\def\eref#1{(\ref{#1})}                                 
\def\lb{\left(}                                         
\def\rb{\right)}                                        
\def\lbs{\left[}                                        
\def\rbs{\right]}                                       
\def\p{\partial}                                        
\def\imag{\mathit{i}}                                   

\begin{document}

\title{Transient and periodic shear wave propagation in a solid-fluid coupled system}
\author{Aaron D'Cruz$^{1}$ and Pierre Ricco$^{1}$}
\address{$^{1}$School of Mechanical, Aerospace and Civil Engineering, The University of Sheffield, Sheffield S1 3JD, UK}
\subject{}
\keywords{transverse waves, damped resonance, transient dynamics, Stokes layer, viscometry}
\corres{Aaron D'Cruz \\ \email{adcruz1413@gmail.com}}

\begin{abstract}
A coupled system composed of a Newtonian fluid located on a sinusoidally-forced elastic solid is studied analytically and numerically. The focus is on the transient evolution from the beginning of the forced oscillations and on the periodic behaviour established once the transient has vanished. The analytical solution is expressed as series summations that elucidate the propagation and reflections of elastic transverse waves through the solid layer and the viscous dissipation of oscillations in the fluid layer. Short-term transients in both the fluid and the solid form at every interaction between an elastic wave and a solid boundary. The long-term transient, quantified by the power balance in the fluid layer, instead pertains to the formation of all the elastic waves in the solid layer. The system can be viewed as a generalised transient Stokes layer generated by the elastic waves or as a damped resonant oscillator when the velocity at the fluid-solid interface increases significantly with respect to the forcing amplitude. A parametric study is carried out for three applications of technological interest, i.e. the indirect measurement of fluid viscosity, the turbulent drag reduction by travelling shear waves and the sensing and manipulation of biological flows.
\end{abstract}

\begin{fmtext}    
\section{Introduction}
\label{introduction}
Shear-driven fluid systems are found in a wide range of engineering and industrial applications. Shear waves, with or without the presence of a bulk flow, may be generated by piezoelectric transducers or electro-osmosis and have been used for sensing, fluid mixing and flow control, particularly at microfluidic scales \cite{ricco-hicks-2018}.
\end{fmtext}

\maketitle

\input{include/1-introduction}
\input{include/2-mathematical-formulation}
\input{include/3-analytical-results}
\input{include/4-numerical-results}
\input{include/5-physical-results}
\input{include/6-conclusions}

\ack{This work was supported by the Engineering and Physical Sciences Research Council (grant number EP/T517835/1) and the Taiho Kogyo Tribology Research Foundation (grant number 191008). The authors are also grateful to Prof. Rob Dwyer-Joyce, Gladys Peretti, Dr Artur Gower, Dr Isabella Fumarola and Prof. Matthew Santer for contributions and informal discussions, to Dr Elena Marensi and Dr Bo Yuan for their constructive feedback, and to Dr Dean Duffy for informative personal correspondence. For the purpose of open access, the author has applied a Creative Commons Attribution (CC BY) licence to any Author Accepted Manuscript version arising.}

\appendix

\input{include/A-contour-integration}
\input{include/B-finite-difference}

\bibliography{pr}{}
\bibliographystyle{unsrt}

\end{document}

%% file: include/1-introduction.tex
Sensing and manipulation of proteins in biochemical flows can be performed via surface acoustic waves (SAWs) with frequencies of the order of GHz, travelling in piezoelectric substrates immersed in a fluid. The interaction between molecules immobilised on the substrate and those suspended in the fluid results in a measurable modification of the SAWs and in an indirect detection of the suspended molecules \cite{lange-etal-2008}. On other SAW devices, standing waves have been used to transport proteins over a few millimetres \cite{neumann-etal-2010}.

Whilst other types of SAWs, such as Rayleigh waves, have also been used in biochemical flows, pure-horizontal waves are favoured due to the lack of attenuation. The strong interaction between the waves and the bulk fluid may be desirable to drive recirculation effects and favour mixing, concentration and separation in microfluidic ``lab-on-a-chip'' devices \cite{shilton-etal-2008,drinkwater-2016}. Ultrasound shear waves are routinely used for laboratory and in-situ monitoring of tribological systems. As lubricating oils in industrial machines experience shear rates as high as $10^7$ $\text{s}^{-1}$, well beyond the range of conventional viscometers, measurements of reflected ultrasonic shear waves are used to investigate the shear rate and temperature dependence of these oils \cite{mason-etal-1949,schirru-etal-2019}. Due to the impedance mismatch between the tested oil and the adjacent solid, a thin matching layer is sometimes used to improve sensitivity \cite{schirru-etal-2015,manfredi-etal-2019}. Shear-wave methods have also been used to measure the mechanical properties of soft tissues  and stresses in railroad steel \cite{li-cao-2017,li-gower-destrade-2020}.

Wall-shear waves have also been utilised to alter turbulent boundary layers and achieve friction drag reduction \cite{choi-2002,quadrio-ricco-2011,marusic-etal-2021}. The modification of wall-bounded turbulent flows by surface waves has been reported to reduce skin friction up to 45\%, although the wall-shear forcing must be specified carefully to avoid drag increase \cite{quadrio-ricco-viotti-2009,bird-santer-morrison-2018,fumarola-etal-2024}.

Some of these applications involve wave transmission through multiple material layers. Mathematical studies of these engineering systems often employ empirical models fitted to experimental data or derived from analogous systems, which necessitate assumptions about the material layers and the geometry of the apparatus \cite{oneil-1949,kasolang-etal-2011}. Some of these models, although derived from simple definitions, become lengthy and complicated for even a few layers of different materials. Such empirical and analogous models are unlikely to be efficient for complex geometries consisting of many layers.

The fundamental mathematical treatment of shear wave propagation through multiple layers is typically focused on the periodic behaviour \cite{goodrich-1969,simonetti-cawley-2004,guo-sun-2008,kielczynski-etal-2012}. In several applications, however, the dynamics over very short timescales is of interest, particularly for real-time monitoring of machinery and for materials with long relaxation times. The initial transient response of a single shear-driven fluid layer to imposed shear wall motion has been investigated \cite{panton-1968}, the motion eventually developing into the classical periodic ``Stokes layer'' \cite{stokes-1851}. More recently, the initial transient response has been studied for a system of two fluid layers with different viscosities \cite{uddin-murad-2022}. Transient shear wave propagation through multiple solid layers is well documented \cite{tang-ting-1985,lee-ma-2000}, but existing transient studies of shear-driven fluids have not considered interactions with adjacent solid layers. Additionally, these studies only describe the initial transient behaviour due to the start of shear forcing, whereas in applications where the forcing is pulsed over short intervals the transient dynamics following the end of the forcing should also be considered.

To the best of our knowledge, a fundamental study that considers the transient response of a coupled solid-fluid system at both the start and end of an interval of shear forcing has not yet been carried out. We therefore present an analytical and numerical investigation of a shear-driven solid layer underneath an unbounded fluid. Although our system is idealised in its geometry, the inclusion of a solid layer through which the fluid is indirectly driven renders this configuration representative of several real setups. We solve the system analytically, arriving at closed-form solutions for the transient and periodic motions of the two layers. The analytical approach leads to insight into the underlying physics that would not be possible with a purely numerical study.

Since our focus is on the physics of the solid-fluid interaction and on technological applications where the viscous penetration depth is thin compared to the distance to the solid boundaries confining the fluid, we consider the fluid to be bounded by the moving boundary only. In other problems where a fluid is forced in oscillatory motion, the fluid is confined between two plates, so that no-slip boundary conditions are imposed at both solid boundaries. The solution to that confined-fluid problem would simplify to our unbounded-fluid solution in the limit of large distance between the solid boundaries \cite{uddin-murad-2022,khaled-vafai-2004}.

In \S\ref{mathematical-formulation} we present the problem in mathematical form using the elastic continuum equation and the Navier-Stokes equation. In \S\ref{analytical-results} we solve the system to find closed-form solutions for the transient and periodic problems and in \S\ref{numerical-results} we solve the equations numerically to verify the accuracy of the analytical solution. In \S\ref{physical-results} we discuss the physics that can be extracted from the analytical solution and explore the behaviour of the system for a range of experimentally inspired parameters. We also quantify the timescales over which the transient phenomena evolve into the periodic motion.

%% file: include/2-mathematical-formulation.tex
\vspace{-2mm}
\section{Mathematical formulation}
\label{mathematical-formulation}

\begin{figure}[h]
    \begin{center}
        \includegraphics[width=.8\textwidth]{"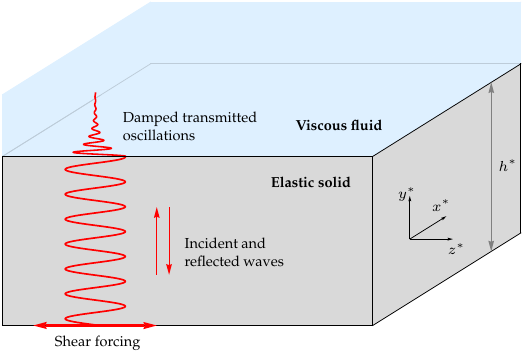"}
        \vspace{-3mm}
        \caption{Schematic of the solid-fluid system.
        \label{geometry}}
    \end{center}
\end{figure}
\vspace{-8mm}
We consider a two-layer system described by Cartesian spatial coordinates $\dimn{x}$, $\dimn{y}$, $\dimn{z}$ and time $\dimn{t}$, where $\dimn{}$ denotes dimensional quantities. A homogeneous, isotropic, elastic solid with density $\dimn{\rho_s}$, Young's modulus $\dimn{E}$ and Poisson ratio $\sigma$ exists between $\dimn{y}=0$ and $\dimn{y}=\dimn{h}$. An incompressible Newtonian fluid with constant kinematic viscosity $\dimn{\nu}$ and density $\dimn{\rho_f}$ sits on top of the solid layer and is unbounded for $\dimn{y}>\dimn{h}$. This system is shown in figure \ref{geometry}. The motion of the solid is described by the elastic continuum equation:

\eqn{
    \dimn{\rho_s} \dfrac{\p^2 \dimn{\boldsymbol{\varphi}}}{\p {\dimn{t}}^2} = \lb \dimn{\Lambda} + \dimn{\mu_s} \rb \nabla \lb \nabla \cdot \dimn{\boldsymbol{\varphi}} \rb + \dimn{\mu_s} \nabla^2 \dimn{\boldsymbol{\varphi}}
\label{3d-solid},
}
\begin{sloppypar}
where $\dimn{\boldsymbol{\varphi}}\lb \dimn{x},\dimn{y},\dimn{z},\dimn{t} \rb$ is the solid displacement and $\dimn{\Lambda}=\dimn{E}\sigma /\lb1+\sigma\rb\lb1-2\sigma\rb$ and $\dimn{\mu_s}=\dimn{E}/2\lb 1+\sigma\rb$ are the Lam\'e parameters. The velocity $\dimn{\vect{u}}\lb \dimn{x},\dimn{y},\dimn{z},\dimn{t} \rb$ of the fluid is described by the Navier-Stokes equation:
\end{sloppypar}

\eqn{
    \dfrac{\p \dimn{\vect{u}}}{\p t} + \lb \dimn{\vect{u}} \cdot \nabla \rb \dimn{\vect{u}} = -\dfrac{\nabla \dimn{p}}{\dimn{\rho_f}} + \dimn{\nu} \nabla^2 \dimn{\vect{u}}
    \label{3d-navierstokes},
}
where $\dimn{p}$ is the pressure. The system is uniform in the $\dimn{x}$ and $\dimn{z}$ directions, and driven by an imposed shear displacement along the $\dimn{z}$ direction at the lower boundary $\dimn{y}=0$. As the solid displacement varies only along a direction normal to the shear deformation, the divergence $\dimn{\nabla\cdot\boldsymbol{\varphi}}$ is zero. {We assume that} there is no pressure gradient in the fluid. Equations \eref{3d-solid} and \eref{3d-navierstokes} thus reduce to a pair of partial differential equations for the shear solid displacement $\dimn{\varphi}$ and the shear fluid velocity $\dimn{u}$ along $\dimn{z}$, as functions of the normal coordinate $\dimn{y}$ and time $\dimn{t}$:

\begin{align}
    \dimn{\rho_s} \dfrac{\p^2 \dimn{\varphi}}{\p {\dimn{t}}^2} = \dimn{\mu_s} \dfrac{\p^2 \dimn{\varphi}}{\p {\dimn{y}}^2}
    \label{eq1*},\\
    \dfrac{\p \dimn{u}}{\p \dimn{t}} = \dimn{\nu} \dfrac{\p^2 \dimn{u}}{\p {\dimn{y}}^2}
    \label{eq2*}.
\end{align}

The boundary conditions are obtained for the forcing of the solid,
\eqn{
    \dimn{\varphi} \lb \dimn{y}=0 \rb = \left\{
        \begin{array}{ccc}
            \dimn{\varphi_0} \sin \lb \dfrac{2 \pi \dimn{t}}{\dimn{T_0}}\rb & \qquad & 0 \leq \dimn{t} < \dimn{\tau}, \\
            0 & & \dimn{t} \geq \dimn{\tau},
        \end{array}\right.
    \label{bc1*}
}
the continuity of shear velocity and shear stress at the solid-fluid interface,
\begin{align}
    \dimn{u} \lb \dimn{y} = \dimn{h} \rb = \left. \dfrac{\p \dimn{\varphi}}{\p \dimn{t}} \right\rvert_{\dimn{y} = \dimn{h}}
    \label{bc2*},\\
    \left. \dimn{\mu_f} \dfrac{\p \dimn{u}}{\p \dimn{y}} \right\rvert_{\dimn{y} = \dimn{h}} = \dimn{\mu_s} \left. \dfrac{\p \dimn{\varphi}}{\p \dimn{y}} \right\rvert_{\dimn{y} = \dimn{h}}
    \label{bc3*},
\end{align}
(where $\dimn{\mu_f}$ is the dynamic viscosity of the fluid) and the vanishing fluid velocity far from the interface,
\eqn{\lim_{\dimn{y}\to\infty} \dimn{u} = 0\label{bc4*}.} We impose stationary initial conditions for the solid and the fluid:
\begin{equation}
    \dimn{\varphi}\lb \dimn{t}=0 \rb=0, \quad
    \left. \dfrac{\p \dimn{\varphi}}{\p \dimn{t}} \right\rvert_{\dimn{t} = 0}=0, \quad
    \dimn{u}\lb \dimn{t}=0 \rb=0.
    \label{ic3*}
\end{equation}

In nondimensional form, the system reads:

\begin{align}
    \dfrac{\p^2 \varphi}{\p t^2} = \dfrac{\p^2 \varphi}{\p {y_{\lambda}}^2}
    \label{eq1}, \\
    \dfrac{\p u}{\p t} = \dfrac{\p^2 u}{\p {y_{\delta}}^2}
    \label{eq2}, \\
    \varphi \lb y_{\lambda} = 0 \rb = \left\{
        \begin{array}{ccc}
            \sin \lb \omega t \rb & \qquad & 0 \leq t < \tau, \\
            0 & & t\geq \tau,
        \end{array}\right.
    \label{bc1} \\
  u \lb y_{\delta} = h_{\delta} \rb = \left. \dfrac{\p \varphi}{\p t} \right\rvert_{y_{\lambda} =  h_{\lambda}}
  \label{bc2}, \\
  \left. \dfrac{\p u}{\p y_{\delta}} \right\rvert_{y_{\delta} = h_{\delta}} = \rho L \left. \dfrac{\p \varphi}{\p y_{\lambda}} \right\rvert_{y_{\lambda} =  h_{\lambda}}
  \label{bc3}, \\
    \lim_{y_{\delta}\to\infty} u = 0
    \label{bc4}, \\
    \varphi\lb t=0 \rb=0, \quad
    \left. \dfrac{\p \varphi}{\p t} \right\rvert_{t = 0}=0, \quad
    u\lb t=0 \rb=0
    \label{ic3},
\end{align}
in terms of the quantities given in table 1. The coordinate $\dimn{y}$ is scaled differently in the two layers. In the solid, $\dimn{y}$ is scaled by the transverse elastic wavelength $\dimn{\lambda} = \dimn{T_0} \sqrt{\dimn{\mu_s} / \dimn{\rho_s}}$ \cite{landau-lifshitz-2012}. In the fluid, $\dimn{y}$ is scaled by the thickness $\dimn{\delta} = \sqrt{\dimn{\nu} \dimn{T_0}}$ of the Stokes layer generated by a sinusoidal wall motion below a still fluid \cite{stokes-1851} since we expect the viscous effects in our case to penetrate to a comparable distance from the solid-fluid interface. The assumption of an unbounded fluid leading to \eref{bc4} is valid if the thickness of the fluid layer is much larger than this Stokes thickness.

\begin{table}[h!]
    \begin{center}
    
    \begin{tabular}{|l|c|}
        \hline
        shear solid displacement & $\varphi$ = $\dimn{\varphi}/\dimn{\varphi_0}$ \\
        \hline
        shear fluid velocity & $u$ = $\dimn{u}\dimn{T_0}/\dimn{\varphi_0}$ \\
        \hline
        time & $t$ = $\dimn{t}/\dimn{T_0}$ \\
        \hline
        angular frequency of sinusoidal forcing & $\omega$ = $2\pi$ \\
        \hline
        duration of sinusoidal forcing & $\tau$ = $\dimn{\tau}/\dimn{T_0}$ \\
        \hline
        wall-normal coordinate in solid & $y_{\lambda}$ = $\dimn{y}/\dimn{\lambda}$ \\
        \hline
        maximum location in solid & $h_{\lambda}$ = $\dimn{h}/\dimn{\lambda}$ \\
        \hline
        wall-normal coordinate in fluid & $y_{\delta}$ = $\dimn{y}/\dimn{\delta}$ \\
        \hline
        minimum location in fluid  & $h_{\delta}$ = $\dimn{h}/\dimn{\delta}$ \\
        \hline
        ratio of densities & $\rho$ = $\dimn{\rho_s}/\dimn{\rho_f}$ \\
        \hline
        ratio of length scales & $L$ = $\dimn{\lambda}/\dimn{\delta}$ \\
        \hline
    \end{tabular}
    \caption{Scaled variables of the coupled system.}
    \end{center}
\end{table}

%% file: include/3-analytical-results.tex
\section{Analytical results}
\label{analytical-results}

We first solve \eref{eq1} - \eref{ic3} analytically in \S \ref{section-laplace} using Laplace transforms to obtain the transient evolution of the two-layer system. At large times, the motion becomes periodic. The solution in this case is obtained in \S\ref{section-fourier} using Fourier modes. In \S\ref{section-fluid-only}, we find the solution for the fluid motion without the solid layer underneath.

\subsection{Transient solution by Laplace transforms}
\label{section-laplace}

The initial value problem \eref{eq1} - \eref{ic3} is solved by using the Laplace transform $\mathcal{L}: t\to s$:
\eqn{\hat{q}\lb s\rb = \mathcal{L} \lbs q \lb t\rb\rbs = \int_0^\infty q\lb t\rb e^{- st} \mathrm{d}t\label{laplace-definition-solid}.}
The system \eref{eq1} - \eref{ic3} is reduced to a system of ordinary differential equations (ODEs) in $\hat{\varphi}$ and $\hat{u}$:
\begin{align}
   s^2 \hat{\varphi} = \dfrac{\mathrm{d}^2 \hat{\varphi}}{\mathrm{d} {y_{\lambda}}^2}
   \label{eq1-laplace}, \\
   s \hat{u} = \dfrac{\mathrm{d}^2 \hat{u}}{\mathrm{d} {y_{\delta}}^2}
   \label{eq2-laplace}, \\
  \hat{\varphi} \lb y_{\lambda} = 0 \rb = \dfrac{\omega\lb1-e^{-{\tau}  s}\lb\cos{\lb\omega \tau \rb}+ \dfrac{s}{\omega}\sin{\lb\omega \tau\rb}\rb\rb}{ s^2+\omega^2}
  \label{bc1-laplace}, \\
  \hat{u} \lb y_{\delta} = h_{\delta} \rb =  s \hat{\varphi} \lb y_{\lambda} =  h_{\lambda} \rb
  \label{bc2-laplace}, \\
  \left. \dfrac{\mathrm{d} \hat{u}}{\mathrm{d} y_{\delta}} \right\rvert_{y_{\delta} = h_{\delta}} = \rho L \left. \dfrac{\mathrm{d} \hat{\varphi}}{\mathrm{d} y_{\lambda}} \right\rvert_{y_{\lambda} =  h_{\lambda}}
  \label{bc3-laplace}, \\
  \lim_{y_{\delta}\to\infty} \hat{u} = 0
  \label{bc4-laplace}.
\end{align}
The solution to \eref{eq1-laplace} - \eref{bc4-laplace} is:
\begin{align}
  \hat{\varphi} = \dfrac{\omega\lb\rho L \cosh{\lb\lb  h_{\lambda}-y_{\lambda}\rb s\rb} +  \sqrt{s} \sinh{\lb\lb  h_{\lambda}-y_{\lambda}\rb s\rb} \rb \hat{T}\lb s \rb}{\lb s^2+\omega^2\rb\lb \rho L \cosh{\lb  h_{\lambda}  s\rb}+ \sqrt{s}\sinh{\lb  h_{\lambda}  s\rb}\rb}
  \label{soln-solid-laplace}, \\
  \hat{u} = \dfrac{\rho L  s \omega e^{\sqrt{s}\lb h_{\delta}-y_{\delta}\rb}\hat{T}\lb s \rb}{\lb s^2+\omega^2\rb\lb\rho L \cosh{\lb  h_{\lambda}  s\rb}+ \sqrt{s}\sinh{\lb  h_{\lambda}  s\rb}\rb}
  \label{soln-fluid-laplace},
\end{align}
where we have defined

\eq{\hat{T}\lb s \rb = 1-e^{-{\tau}  s}\lb\cos{\lb\omega \tau \rb}+ \dfrac{s}{\omega}\sin{\lb\omega \tau\rb}\rb.}
The exponentially growing term arising in the solution to \eqref{eq2-laplace} is zero due to the fluid being unbounded, as given by \eref{bc4}. If the fluid were confined, this exponential term would be non-zero and the fluid-layer solution would involve hyperbolic functions at the numerator, as in the solid-layer solution \eqref{soln-solid-laplace}. In order to recover the physical solutions $\varphi \lb y_{\lambda}, t \rb $ and $u \lb y_{\delta}, t \rb$, the inverse Laplace transform $\mathcal{L}^{-1}$ is applied. First, we write the hyperbolic functions in \eref{soln-solid-laplace} and \eref{soln-fluid-laplace} as exponentials and divide the numerator and denominator by $e^{h_{\lambda}s}$, obtaining a common denominator expressed as a geometric ray series \cite{duffy-2004}:
\eqn{
  \hat{B}\lb s \rb={\lb 1+ \dfrac{\rho L- \sqrt{s}}{\rho L+ \sqrt{s}} e^{- 2  h_{\lambda}  s} \rb}^{-1}=\sum_{n=0}^{\infty} \lb -1\rb^n \lb\dfrac{\rho L- \sqrt{s}}{\rho L+ \sqrt{s}}\rb^n e^{-2n h_{\lambda} s}.\label{B-definintion}
}

The convergence of the series in \eref{B-definintion} is justified since $ h_{\lambda}>0$ and $\mathrm{Re}\lbs s\rbs>0$ for the inverse Laplace transform. We obtain
\begin{align}
  \hat{\varphi} &= \dfrac{e^{- s y_{\lambda}}\omega\hat{T}\lb s \rb\hat{B}\lb s \rb}{s^2+\omega^2}+ \dfrac{e^{s\lb y_{\lambda}-2 h_{\lambda}\rb}\lb\rho L- \sqrt{s}\rb\omega\hat{T}\lb s \rb\hat{B}\lb s \rb}{\lb s^2+\omega^2 \rb\lb\rho L+ \sqrt{s}\rb}
  \label{laplace-B-soln-solid}, \\
  \hat{u} &= \dfrac{2 \rho L  s e^{\sqrt{s}\lb h_{\delta}-y_{\delta}\rb- s h_{\lambda}}\omega\hat{T}\lb s \rb\hat{B}\lb s \rb}{\lb s^2+\omega^2 \rb\lb\rho L+ \sqrt{s}\rb}
  \label{laplace-B-soln-fluid}.
\end{align}

Since $0\leq y_{\lambda}\leq  h_{\lambda}$, we use the time shifting property of the Laplace transform to replace the exponents involving $-s$ in \eref{laplace-B-soln-solid} and \eref{laplace-B-soln-fluid} with Heaviside step functions $\mathrm{H}$:
\begin{align*}
  \varphi \lb y_{\lambda},t\rb =& \mathcal{L}^{-1} \lbs\hat{\varphi}\lb y_{\lambda}, s\rb\rbs = \mathcal{L}^{-1} \left[ \dfrac{e^{- s y_{\lambda}}\omega\hat{T}\lb s \rb\hat{B}\lb s \rb}{s^2+\omega^2}
    + \dfrac{e^{s\lb y_{\lambda}-2 h_{\lambda}\rb}\lb\rho L- \sqrt{s}\rb\omega\hat{T}\lb s \rb\hat{B}\lb s \rb}{\lb s^2+\omega^2 \rb\lb\rho L+ \sqrt{s}\rb} \right] \\
  =& \mathrm{H}\lb t- y_{\lambda}\rb \mathcal{L}^{-1} \lbs \dfrac{\omega\hat{T}\lb s \rb\hat{B}\lb s \rb}{s^2+\omega^2}\rbs_{t \to t-y_{\lambda}} 
  \\
  +&\mathrm{H}\lb t + y_{\lambda}-2 h_{\lambda } \rb \mathcal{L}^{-1} \lbs \dfrac{\lb\rho L- \sqrt{s}\rb\omega\hat{T}\lb s \rb\hat{B}\lb s \rb}{\lb s^2+\omega^2 \rb\lb\rho L+ \sqrt{s}\rb} \rbs_{t \to t+y_{\lambda}-2 h_{\lambda}}
  \label{},\\
  u \lb y_{\delta},t\rb =& \mathcal{L}^{-1} \lbs\hat{u}\lb y_{\delta}, s\rb\rbs = \mathcal{L}^{-1} \lbs \dfrac{2 \rho L  s e^{\sqrt{s}\lb h_{\delta}-y_{\delta}\rb- s h_{\lambda}}\omega\hat{T}\lb s \rb\hat{B}\lb s \rb}{\lb s^2+\omega^2 \rb\lb\rho L+ \sqrt{s}\rb} \rbs \\
  =& \mathrm{H} \lb t- h_{\lambda} \rb \mathcal{L}^{-1} \lbs \dfrac{2 \rho L  s e^{\sqrt{s}\lb h_{\delta}-y_{\delta}\rb}\omega\hat{T}\lb s \rb\hat{B}\lb s \rb}{\lb s^2+\omega^2 \rb\lb\rho L+ \sqrt{s}\rb} \rbs_{t \to t- h_{\lambda}}
  \label{}.
\end{align*}

By further application of the time shifting property, the exponents in $\hat{B}\lb s \rb$ may be replaced using another time-coordinate shift:
\begin{align*}
  \varphi = \sum_{n=0}^{\infty} \lb -1\rb^n &\mathrm{H}\lb t- y_{\lambda} -2n h_{\lambda}\rb \mathcal{L}^{-1} \lbs  \lb\dfrac{\rho L- \sqrt{s}}{\rho L+ \sqrt{s}}\rb^n \dfrac{\omega\hat{T}\lb s\rb}{s^2+\omega^2}\rbs_{t \to t-y_{\lambda}-2nh_{\lambda}} \\
  + \sum_{n=0}^{\infty} \lb -1\rb^n &\mathrm{H}\lb t+ y_{\lambda}-2\lb n+1 \rb h_{\lambda}\rb \mathcal{L}^{-1} \lbs \lb\dfrac{\rho L- \sqrt{s}}{\rho L+ \sqrt{s}}\rb^{n+1} \dfrac{\omega\hat{T}\lb s\rb}{s^2+\omega^2}\rbs_{t \to t+y_{\lambda}-2 \lb n+1\rb h_{\lambda}}
  \label{},\\
  u =  \sum_{n=0}^{\infty} \lb -1\rb^n &\mathrm{H} \lb t- \lb 2n+1\rb h_{\lambda} \rb \mathcal{L}^{-1} \lbs \dfrac{2 \rho L  s \omega\lb\rho L- \sqrt{s}\rb^n e^{\sqrt{s}\lb h_{\delta}-y_{\delta}\rb}\hat{T}\lb s\rb}{\lb s^2+\omega^2 \rb\lb\rho L+ \sqrt{s}\rb^{n+1}} \rbs_{t \to t- \lb 2n+1\rb h_{\lambda}.}
\end{align*}

The introduction of the summations, with each term shifted in time by $2nh_{\lambda}$, clarifies the physical interpretation of the solid and fluid solutions as series of superposed reflections evolving in time. Applying the time shifting property once more to the exponents in $\hat{T}\lb s \rb$, we obtain:
\begin{align*}
  \varphi = \sum_{n=0}^{\infty} \lb -1\rb^n &\mathrm{H}\lb t- y_{\lambda} -2n h_{\lambda}\rb \mathcal{L}^{-1} \lbs  \lb\dfrac{\rho L- \sqrt{s}}{\rho L+ \sqrt{s}}\rb^n \dfrac{\omega}{s^2+\omega^2}\rbs_{t \to t-y_{\lambda}-2nh_{\lambda}} \\
  - \cos{\lb\omega \tau \rb} \sum_{n=0}^{\infty} \lb -1\rb^n &\mathrm{H}\lb t- y_{\lambda} -2n h_{\lambda}-\tau\rb \mathcal{L}^{-1} \lbs  \lb\dfrac{\rho L- \sqrt{s}}{\rho L+ \sqrt{s}}\rb^n \dfrac{\omega}{s^2+\omega^2}\rbs_{t \to t-y_{\lambda}-2nh_{\lambda}-\tau} \\
  - \sin{\lb\omega \tau \rb} \sum_{n=0}^{\infty} \lb -1\rb^n &\mathrm{H}\lb t- y_{\lambda} -2n h_{\lambda}-\tau\rb \mathcal{L}^{-1} \lbs  \lb\dfrac{\rho L- \sqrt{s}}{\rho L+ \sqrt{s}}\rb^n \dfrac{s}{s^2+\omega^2}\rbs_{t \to t-y_{\lambda}-2nh_{\lambda}-\tau}
\end{align*}
\begin{align*}
  + \sum_{n=0}^{\infty} \lb -1\rb^n &\mathrm{H}\lb t+ y_{\lambda}-2\lb n+1 \rb h_{\lambda}\rb \mathcal{L}^{-1} \lbs \lb\dfrac{\rho L- \sqrt{s}}{\rho L+ \sqrt{s}}\rb^{n+1} \dfrac{\omega}{s^2+\omega^2}\rbs_{\substack{t \to t+y_{\lambda}\qquad\\-2 \lb n+1\rb h_{\lambda}}} \\
  - \cos{\lb\omega \tau \rb} \sum_{n=0}^{\infty} \lb -1\rb^n &\mathrm{H}\lb t+ y_{\lambda}-2\lb n+1 \rb h_{\lambda}-\tau\rb \mathcal{L}^{-1} \lbs \lb\dfrac{\rho L- \sqrt{s}}{\rho L+ \sqrt{s}}\rb^{n+1} \dfrac{\omega}{s^2+\omega^2}\rbs_{\substack{t \to t+y_{\lambda}\qquad\\-2 \lb n+1\rb h_{\lambda}-\tau}} \\
  - \sin{\lb\omega \tau \rb} \sum_{n=0}^{\infty} \lb -1\rb^n &\mathrm{H}\lb t+ y_{\lambda}-2\lb n+1 \rb h_{\lambda}-\tau\rb \mathcal{L}^{-1} \lbs \lb\dfrac{\rho L- \sqrt{s}}{\rho L+ \sqrt{s}}\rb^{n+1} \dfrac{s}{s^2+\omega^2}\rbs_{\substack{t \to t+y_{\lambda}\qquad\\-2 \lb n+1\rb h_{\lambda}-\tau}},
\end{align*}
\begin{align*}
  u =  \sum_{n=0}^{\infty} \lb -1\rb^n &\mathrm{H} \lb t- \lb 2n+1\rb h_{\lambda} \rb \mathcal{L}^{-1} \lbs \dfrac{2 \rho L  s \omega\lb\rho L- \sqrt{s}\rb^n e^{\sqrt{s}\lb h_{\delta}-y_{\delta}\rb}}{\lb s^2+\omega^2 \rb\lb\rho L+ \sqrt{s}\rb^{n+1}} \rbs_{t \to t- \lb 2n+1\rb h_{\lambda}} \\
  -\cos{\lb\omega \tau \rb} \sum_{n=0}^{\infty} \lb -1\rb^n &\mathrm{H} \lb t- \lb 2n+1\rb h_{\lambda} -\tau \rb \mathcal{L}^{-1} \lbs \dfrac{2 \rho L  s \omega\lb\rho L- \sqrt{s}\rb^n e^{\sqrt{s}\lb h_{\delta}-y_{\delta}\rb}}{\lb s^2+\omega^2 \rb\lb\rho L+ \sqrt{s}\rb^{n+1}} \rbs_{t \to t- \lb 2n+1\rb h_{\lambda}-\tau} \\
  -\sin{\lb\omega \tau \rb} \sum_{n=0}^{\infty} \lb -1\rb^n &\mathrm{H} \lb t- \lb 2n+1\rb h_{\lambda} -\tau \rb \mathcal{L}^{-1} \lbs \dfrac{2 \rho L  s^2 \lb\rho L- \sqrt{s}\rb^n e^{\sqrt{s}\lb h_{\delta}-y_{\delta}\rb}}{\lb s^2+\omega^2 \rb\lb\rho L+ \sqrt{s}\rb^{n+1}} \rbs_{t \to t- \lb 2n+1\rb h_{\lambda}-\tau}.
\end{align*}
Although the summation index of the series goes from $0$ to $\infty$, for a large enough $n$ and given $t$ and $y_{\lambda}$, only a finite number of terms are required for the solutions to be exact because of the presence of the Heaviside functions. The remaining inverse Laplace transforms may be grouped and rewritten as families of functions $G_{\mathcal{S},m,n}\lb t\rb$ and $G_{\mathcal{F},m,n}\lb y_{\delta},t\rb$:
\begin{align}
  G_{\mathcal{S},m,n}\lb t\rb=\mathcal{L}^{-1}\lbs \dfrac{ s^m \omega^{1-m}}{s^2+\omega^2 }\lb \dfrac{\rho L - \sqrt{s}}{\rho L + \sqrt{s}}\rb ^n\rbs,
  \label{GSmn-definition}\\
  G_{\mathcal{F},m,n}\lb y_{\delta}, t\rb=\mathcal{L}^{-1}\lbs \dfrac{2 \rho L  s^{1+m} \omega^{1-m} \lb \rho L - \sqrt{s}\rb ^n e^{\sqrt{s}\lb h_{\delta}-y_{\delta}\rb}}{\lb  s^2+\omega^2\rb \lb \rho L + \sqrt{s}\rb ^{n+1}}\rbs
  \label{GFmn-definition},
\end{align}
with integer indices $m=0,1$, and $n\geq 0$. Using these functions, the solutions $\varphi$ and $u$ may be written in terms of $G_{\mathcal{S},0,n}$, $G_{\mathcal{S},1,n}$, $G_{\mathcal{F},0,n}$ and $G_{\mathcal{F},1,n}$:
\begin{align*}
  \varphi=&\sum_{n=0}^\infty \lb -1\rb ^n \mathrm{H}\lb t-y_{\lambda}-2n h_{\lambda}\rb  G_{\mathcal{S},0,n}\lb t-y_{\lambda}-2n h_{\lambda}\rb  \notag\\
  -\cos{\lb \omega{\tau}\rb } &\sum_{n=0}^\infty \lb -1\rb ^n \mathrm{H}\lb t-y_{\lambda}-2n h_{\lambda}-{\tau}\rb  G_{\mathcal{S},0,n}\lb t-y_{\lambda}-2n h_{\lambda}-{\tau}\rb \notag\\
  -\sin{\lb \omega{\tau}\rb } &\sum_{n=0}^\infty \lb -1\rb ^n \mathrm{H}\lb t-y_{\lambda}-2n h_{\lambda}-{\tau}\rb  G_{\mathcal{S},1,n}\lb t-y_{\lambda}-2n h_{\lambda}-{\tau}\rb \notag
\end{align*}
\begin{align}
  +&\sum_{n=0}^\infty \lb -1\rb ^n \mathrm{H}\lb t+y_{\lambda}-2 \lb n+1\rb  h_{\lambda}\rb  G_{\mathcal{S},0,n+1}\lb t+y_{\lambda}-2 \lb n+1\rb  h_{\lambda}\rb \notag \\
  -\cos{\lb \omega{\tau}\rb } &\sum_{n=0}^\infty \lb -1\rb ^n \mathrm{H}\lb t+y_{\lambda}-2 \lb n+1\rb  h_{\lambda}-{\tau}\rb  G_{\mathcal{S},0,n+1}\lb t+y_{\lambda}-2 \lb n+1\rb  h_{\lambda}-{\tau}\rb \notag\\
  -\sin{\lb \omega{\tau}\rb } &\sum_{n=0}^\infty \lb -1\rb ^n \mathrm{H}\lb t+y_{\lambda}-2 \lb n+1\rb  h_{\lambda}-{\tau}\rb  G_{\mathcal{S},1,n+1}\lb t+y_{\lambda}-2 \lb n+1\rb  h_{\lambda}-{\tau}\rb
  \label{transient-solid-soln},
\end{align}
\begin{align}
  u=&\sum_{n=0}^\infty \lb -1\rb ^n \mathrm{H}\lb t-\lb 2n+1\rb  h_{\lambda}\rb  G_{\mathcal{F},0,n}\lb y_{\delta}, t-\lb 2n+1\rb  h_{\lambda}\rb \notag \\
  -\cos{\lb \omega{\tau}\rb } &\sum_{n=0}^\infty \lb -1\rb ^n \mathrm{H}\lb t-\lb 2n+1\rb  h_{\lambda}-{\tau}\rb  G_{\mathcal{F},0,n}\lb y_{\delta},t-\lb 2n+1\rb  h_{\lambda}-{\tau}\rb \notag \\
  -\sin{\lb \omega{\tau}\rb } &\sum_{n=0}^\infty \lb -1\rb ^n \mathrm{H}\lb t-\lb 2n+1\rb  h_{\lambda}-{\tau}\rb  G_{\mathcal{F},1,n}\lb y_{\delta},t-\lb 2n+1\rb  h_{\lambda}-{\tau}\rb \label{transient-fluid-soln}.
\end{align}
$G_{\mathcal{S},0,n}$, $G_{\mathcal{S},1,n}$, $G_{\mathcal{F},0,n}$ and $G_{\mathcal{F},1,n}$ are found by integration in the complex plane along a Bromwich contour (the detailed derivation is given in Appendix \ref{contour-integration}). These functions are expressed in physical space in terms of parameters $\mathcal{K}_{0-7}$:
\eqn{G_{\mathcal{S},0,n}=
\dfrac{\mathcal{K}_1}{\mathcal{K}_0^n}\sin{\lb\omega t\rb}+\dfrac{\mathcal{K}_2}{\mathcal{K}_0^n}\cos{\lb\omega t\rb}
-\dfrac{1}{\pi}\int_0^\infty \mathcal{K}_3 \dfrac{\omega e^{-Rt}}{R^2+\omega^2} \mathrm{d}R,
\label{GS0-exact}
}
\eqn{G_{\mathcal{S},1,n}=
\dfrac{\mathcal{K}_1}{\mathcal{K}_0^n}\cos{\lb\omega t\rb}-\dfrac{\mathcal{K}_2}{\mathcal{K}_0^n}\sin{\lb\omega t\rb}
+\dfrac{1}{\pi}\int_0^\infty \mathcal{K}_3 \dfrac{R e^{-Rt}}{R^2+\omega^2} \mathrm{d}R,
\label{GS1-exact}
}
\eqn{
\begin{split}  
G_{\mathcal{F},0,n}=\dfrac{2\rho L\omega e^{\lb h_{\delta}-y_{\delta} \rb \sqrt{\omega /2}}}{\mathcal{K}_0^{n+1}} &\lb \mathcal{K}_4 \cos{\lb\lb h_{\delta}-y_{\delta} \rb \sqrt{\omega /2}+\omega t\rb} +\mathcal{K}_5 \sin{\lb\lb h_{\delta}-y_{\delta} \rb \sqrt{\omega /2}+\omega t\rb} \rb \\
+\dfrac{1}{\pi}\int_0^\infty \dfrac{2\rho L\omega R e^{-Rt}}{\lb R^2 + \omega^2\rb\lb\rho^2 L^2 +R\rb}&\lb
\mathcal{K}_6\sin{\lb\lb h_{\delta}-y_{\delta} \rb \sqrt{R}\rb}+\mathcal{K}_7\cos{\lb\lb h_{\delta}-y_{\delta} \rb \sqrt{R}\rb}
\rb\mathrm{d}R,
\label{GF0-exact}
\end{split}
}
\eqn{
\begin{split}  
G_{\mathcal{F},1,n}=\dfrac{2\rho L\omega e^{\lb h_{\delta}-y_{\delta} \rb \sqrt{\omega /2}}}{\mathcal{K}_0^{n+1}} &\lb \mathcal{K}_5 \cos{\lb\lb h_{\delta}-y_{\delta} \rb \sqrt{\omega /2}+\omega t\rb} -\mathcal{K}_4 \sin{\lb\lb h_{\delta}-y_{\delta} \rb \sqrt{\omega /2}+\omega t\rb} \rb \\
-\dfrac{1}{\pi}\int_0^\infty \dfrac{2\rho L\omega R^2 e^{-Rt}}{\lb R^2 + \omega^2\rb\lb\rho^2 L^2 +R\rb}&\lb
\mathcal{K}_6\sin{\lb\lb h_{\delta}-y_{\delta} \rb \sqrt{R}\rb}+\mathcal{K}_7\cos{\lb\lb h_{\delta}-y_{\delta} \rb \sqrt{R}\rb}
\rb\mathrm{d}R,
\label{GF1-exact}
\end{split}
}
where
\eq{
\begin{array}{cc}  
\mathcal{K}_0=\lb\rho L+\sqrt{\omega /2}\rb^2+\omega/2, &
\mathcal{K}_1=\mathrm{Re}\lbs\lb \rho^2 L^2-\omega-2\imag\rho L\sqrt{\omega /2} \rb^n\rbs, \\\\
\mathcal{K}_2=\mathrm{Im}\lbs\lb \rho^2 L^2-\omega-2\imag\rho L\sqrt{\omega /2} \rb^n\rbs, &
\mathcal{K}_3\lb R \rb=\mathrm{Im}\lbs \lb\dfrac{\lb\rho L-\imag\sqrt{R}\rb^{2}}{\rho^2 L^2+R}\rb^n \rbs,\\\\
\mathcal{K}_4=\lb\rho L+\sqrt{\omega /2}\rb \mathcal{K}_1+\sqrt{\omega /2} \mathcal{K}_2,&
\mathcal{K}_5=\sqrt{\omega /2}\mathcal{K}_1-\lb\rho L+\sqrt{\omega /2}\rb \mathcal{K}_2,
\end{array}}
\eq{\mathcal{K}_6\lb R\rb=\rho L \mathrm{Re}\lbs \lb\dfrac{\rho L-\imag\sqrt{R}}{\rho L+\imag\sqrt{R}}\rb^n \rbs+\sqrt{R}\mathrm{Im}\lbs \lb\dfrac{\rho L-\imag\sqrt{R}}{\rho L+\imag\sqrt{R}}\rb^n \rbs,}
\eq{\mathcal{K}_7\lb R\rb=\rho L \mathrm{Im}\lbs \lb\dfrac{\rho L-\imag\sqrt{R}}{\rho L+\imag\sqrt{R}}\rb^n \rbs-\sqrt{R}\mathrm{Re}\lbs \lb\dfrac{\rho L-\imag\sqrt{R}}{\rho L+\imag\sqrt{R}}\rb^n \rbs.}

$G_{\mathcal{S},0,n}$, $G_{\mathcal{S},1,n}$, $G_{\mathcal{F},0,n}$ and $G_{\mathcal{F},1,n}$ consist of time-periodic terms involving $\sin{\lb\omega t\rb}$ and $\cos{\lb\omega t\rb}$, plus terms decaying exponentially in time. Two transient contributions to the solutions therefore exist: the summation of terms in \eref{transient-solid-soln} and \eref{transient-fluid-soln} where the number of contributing terms depends on $t$ through the Heaviside functions, and the exponential terms inside $G_{\mathcal{S},0,n}$, $G_{\mathcal{S},1,n}$, $G_{\mathcal{F},0,n}$ and $G_{\mathcal{F},1,n}$. At large times, the system approaches a periodic state given by the full summations in \eref{transient-solid-soln} and \eref{transient-fluid-soln} and the periodic terms in $G_{\mathcal{S},0,n}$, $G_{\mathcal{S},1,n}$, $G_{\mathcal{F},0,n}$ and $G_{\mathcal{F},1,n}$. A detailed discussion on the physical interpretation of the analytical solutions is given in \S \ref{physical-results-analytical}.

\subsection{Periodic solution by Fourier modes}
\label{section-fourier}

As $\tau\to\infty$, the terms involving $\tau$ in the transient solutions \eref{transient-solid-soln} and \eref{transient-fluid-soln} vanish due to their multiplying Heaviside functions having negative arguments for all $t\geq 0$. In this simplification, which may also be obtained by setting $\hat{T}\lb s\rb=1$ in \eref{soln-solid-laplace} and \eref{soln-fluid-laplace}, the forcing in \eref{bc1} is sinusoidal for $t\geq 0$. Furthermore, for sufficiently large values of $t$, the motion becomes independent from the initial conditions and the expressions \eref{transient-solid-soln} and \eref{transient-fluid-soln} become periodic in time, i.e. when all the transient phenomena have decayed. Therefore, assuming time-periodic solutions:

\begin{align}
  \varphi_{\mathrm{P}}\lb y_{\lambda},t\rb = \mathrm{Im}\lbs \tilde{\varphi}\lb y_{\lambda}\rb e^{\imag\omega t}\rbs, \quad
  u_{\mathrm{P}}\lb y_{\delta},t\rb = \mathrm{Im}\lbs \tilde{u}\lb y_{\delta}\rb e^{\imag\omega t}\rbs,
\end{align}
equations \eref{eq1} - \eref{bc4} are rewritten as a system of ODEs for $\tilde{\varphi}$ and $\tilde{u}$:
\begin{align}
  \imag\omega \tilde{u} = \dfrac{\mathrm{d}^2 \tilde{u}}{\mathrm{d} {y_{\delta}}^2}
  \label{eq1-fourier}, \\
  -\omega^2 \tilde{\varphi} = \dfrac{\mathrm{d}^2 \tilde{\varphi}}{\mathrm{d} {y_{\lambda}}^2}
  \label{eq2-fourier},\\
  \tilde{\varphi} \lb y_{\lambda} = 0 \rb = 1
  \label{bc1-fourier},\\
  \tilde{u} \lb y_{\delta} = h_{\delta} \rb = \left. \imag\omega \tilde{\varphi} \right\rvert_{y_{\lambda} =  h_{\lambda}}
  \label{bc2-fourier},\\
  \left. \dfrac{\mathrm{d} \tilde{u}}{\mathrm{d} y_{\delta}} \right\rvert_{y_{\delta} = h_{\delta}} = \rho L \left. \dfrac{\mathrm{d} \tilde{\varphi}}{\mathrm{d} y_{\lambda}} \right\rvert_{y_{\lambda} =  h_{\lambda}}
  \label{bc3-fourier},\\
  \lim_{y_{\delta}\to\infty} \tilde{u} = 0
  \label{bc4-fourier}.
\end{align}
The solution to \eref{eq1-fourier} - \eref{bc4-fourier} is:

\begin{align}
  \tilde{\varphi}=\dfrac{\rho L \cos{\lb\omega \lb  h_{\lambda}-y_{\lambda}\rb\rb}+\imag \sqrt{\imag\omega}\sin{\lb\omega \lb  h_{\lambda}-y_{\lambda}\rb\rb}}{\rho L \cos{\lb\omega  h_{\lambda}\rb}+\imag\sqrt{\imag\omega}\sin{\lb\omega  h_{\lambda}\rb}},\\
  \tilde{u}=\dfrac{\imag\rho L\omega e^{\lb h_{\delta}-y_{\delta}\rb\sqrt{\imag\omega}}}{\rho L \cos{\lb\omega  h_{\lambda}\rb}+\imag\sqrt{\imag\omega}\sin{\lb\omega h_{\lambda}\rb}}.
\end{align}
The solutions $\varphi_{\mathrm{P}}$ and $u_{\mathrm{P}}$ are expressed in terms of parameters $\mathcal{P}_{1-5}$:
\begin{align}
  \varphi_{\mathrm{P}} = \mathrm{Im}\lbs \tilde{\varphi} e^{\imag\omega t}\rbs=&\dfrac{\mathcal{P}_1\lb y_{\lambda}\rb}{\mathcal{P}_5} \sin{\lb \omega t \rb} + \dfrac{\mathcal{P}_2\lb y_{\lambda}\rb}{\mathcal{P}_5} \cos{\lb \omega t \rb}
  \label{periodic-solid-soln},\\
  u_{\mathrm{P}} = \mathrm{Im}\lbs \tilde{u} e^{\imag\omega t}\rbs = &\dfrac{\mathcal{P}_3}{\mathcal{P}_5}e^{\lb h_{\delta}-y_{\delta}\rb\sqrt{\omega/2}}\sin{\lb\lb h_{\delta}-y_{\delta}\rb\sqrt{\omega/2}+\omega t\rb} \label{periodic-fluid-soln} \notag\\
  +&\dfrac{\mathcal{P}_4}{\mathcal{P}_5}e^{\lb h_{\delta}-y_{\delta}\rb\sqrt{\omega/2}}\cos{\lb\lb h_{\delta}-y_{\delta}\rb\sqrt{\omega/2}+\omega t\rb},
\end{align}
where
\vspace{-3mm}
\eq{
  \begin{split}
  &\begin{split}
  \mathcal{P}_1\lb y_{\lambda}\rb
  =
  &
  \rho^2 L^2 \cos{\lb\omega h_{\lambda}\rb}\cos{\lb\omega\lb h_{\lambda} - y_\lambda \rb\rb} -\rho L \sqrt{\dfrac{\omega}{2}}\left[\sin{\lb\omega h_{\lambda}\rb}\cos{\lb\omega\lb h_{\lambda} - y_\lambda \rb\rb}
  \right.\\
  & \left.
  +\cos{\lb\omega h_{\lambda}\rb}\sin{\lb\omega\lb h_{\lambda} - y_\lambda \rb\rb}
  \right]
  +\omega \sin{\lb\omega h_{\lambda}\rb}\sin{\lb\omega\lb h_{\lambda} - y_\lambda \rb\rb},
  \end{split}\\
    &\mathcal{P}_2\lb y_{\lambda}\rb= \rho L\sqrt{\dfrac{\omega}{2}}\lb\cos{\lb\omega h_{\lambda}\rb}\sin{\lb\omega \lb h_{\lambda}-y_{\lambda} \rb\rb}-\sin{\lb\omega h_{\lambda}\rb}\cos{\lb\omega \lb h_{\lambda}-y_{\lambda} \rb\rb}\rb, \\
    &\mathcal{P}_3= \rho L \omega \sqrt{\omega/2} \sin{\lb \omega h_{\lambda} \rb}, \quad
    \mathcal{P}_4= \rho L \omega \lb \rho L \cos{\lb \omega h_{\lambda} \rb} - \sqrt{\omega/2} \sin{\lb \omega h_{\lambda} \rb} \rb, \\
    &\mathcal{P}_5= \lb \rho L \cos{\lb \omega h_{\lambda} \rb} - \sqrt{\omega/2} \sin{\lb \omega h_{\lambda} \rb} \rb^2 + \lb \sqrt{\omega/2} \sin{\lb \omega h_{\lambda} \rb} \rb^2.
  \end{split}
}
The two terms in $u_{\mathrm{P}}$ can be combined by assuming a solution of the form:
\eqn{
  u_{\mathrm{P}} = \mathcal{A}e^{\lb h_{\delta}-y_{\delta}\rb\sqrt{\omega/2}}\sin{\lb\lb h_{\delta}-y_{\delta}\rb\sqrt{\omega/2}+\omega \lb t+\mathcal{T}\rb\rb}
  \label{periodic-fluid-soln-merged},
}
and then expanding and matching coefficients with those in \eref{periodic-fluid-soln}. The periodic fluid solution \eref{periodic-fluid-soln-merged} is in the form of a Stokes layer with a time offset $\mathcal{T}$ and amplitude $\mathcal{A}$:
\eqn{
  \mathcal{T}=\dfrac{1}{\omega}\arctan{\lb\sqrt{\dfrac{2}{\omega}}\rho L \cot{\lb \omega h_{\lambda} \rb}-1\rb} \qquad \mathcal{A}=\dfrac{\mathcal{P}_3}{\mathcal{P}_5}\sec{\lb\omega\mathcal{T}\rb}.
  \label{periodic-fluid-soln-merged-amplitude}
}

The maximum velocity $\mathcal{A}$ is the same as the maximum velocity in the solid at $y_{\lambda}=h_{\lambda}$, obtained by differentiating \eref{periodic-solid-soln} with respect to time, and depends on $h_{\lambda}$ and $\rho L$. The periodic motion of the fluid in the two-layer system therefore has the same functional form as that of a fluid driven directly at its lower boundary, with the solid layer underneath serving only to vary the amplitude and phase of the Stokes layer. The value of $\mathcal{A}$ can can be positive or negative due to the sign of $\mathcal{P}_3$, so the overall phase difference between the material interface and the lower boundary of the solid depends on both $\mathcal{T}$ and the sign of $\mathcal{A}$.

\subsection{Fluid-only solution}
\label{section-fluid-only}

We now consider the case of a vanishingly small solid layer, where the fluid is forced directly at its lower boundary. A simplified Laplace-space solution $\hat{u}_\mathcal{F}$ is obtained for the fluid by setting $h_{\lambda}=h_{\delta}=0$ in \eref{soln-fluid-laplace}:

\eqn{\hat{u}_\mathcal{F}=\dfrac{s \omega e^{-y_{\delta} \sqrt{s}}\lb 1-e^{-{\tau}  s}\lb\cos{\lb\omega \tau \rb}+ \dfrac{s}{\omega}\sin{\lb\omega \tau\rb}\rb\rb}{s^2+\omega^2}.
\label{fluid-only-laplace}
}

Applying the inverse Laplace transform to \eref{fluid-only-laplace} and using the same method given in Appendix \ref{contour-integration} for $G_{\mathcal{S},m,n}$ and $G_{\mathcal{F},m,n}$, we obtain the corresponding solution in physical space:
\begin{align}
u_\mathcal{F}=\omega e^{-y_{\delta}\sqrt{\omega/2}} \cos{\lb \omega t - y_{\delta}\sqrt{\omega/2} \rb}&-\dfrac{\omega}{\pi} \int_0^\infty \dfrac{R}{R^2+\omega^2} e^{-Rt} \sin{\lb y_{\delta} \sqrt{R} \rb} \mathrm{d}R \notag \\
-\mathrm{H}\lbs t-\tau \rbs
\cos{\lb\omega \tau\rb} &\left( \omega e^{-y_{\delta}\sqrt{\omega/2}} \cos{\lb \omega \lb t-\tau \rb - y_{\delta}\sqrt{\omega/2} \rb} \right.\notag\\
&\left.-\dfrac{\omega}{\pi} \int_0^\infty \dfrac{R}{R^2+\omega^2} e^{-R\lb t-\tau\rb} \sin{\lb y_{\delta} \sqrt{R} \rb} \mathrm{d}R \right) \notag\\
-\mathrm{H}\lbs t-\tau \rbs
\sin{\lb\omega \tau\rb} &\left( e^{-y_{\delta}\sqrt{\omega/2}} \sin{\lb \omega \lb t-\tau\rb - y_{\delta}\sqrt{\omega/2} \rb}\right.\notag\\
&
\left.+\dfrac{1}{\pi} \int_0^\infty \dfrac{R^2}{R^2+\omega^2} e^{-R\lb t-\tau\rb} \sin{\lb y_{\delta} \sqrt{R} \rb} \mathrm{d}R \right) \label{fluid-only-soln}.
\end{align}

In the limit $\tau\to\infty$, only the first two terms in \eref{fluid-only-soln} are non-zero. The solution matches the solution to the transient extension of the Stokes second problem \cite{panton-1968} and the equivalent heat transfer problem \cite{carslaw-jaeger-1959}. Similarly, setting $h_{\lambda}=h_{\delta}=0$ in \eref{periodic-fluid-soln} leads to the first term of \eref{fluid-only-soln}, i.e. the periodic Stokes layer \cite{stokes-1851}. The solution to the two-layer system therefore reduces as expected to that of a single layer of viscous fluid in both the transient and periodic cases.

%% file: include/4-numerical-results.tex
\section{Numerical procedures}
\label{numerical-results}

We solve the system \eref{eq1} - \eref{ic3} numerically to verify the accuracy of the analytical solution. We adapt the method of Cebeci \cite{cebeci-bradshaw-1984} to allow for the interface conditions \eref{bc2} and \eref{bc3}. The finite-difference discretisation of the governing equations is performed by first reducing the second-order spatial derivatives in \eref{eq1} and \eref{eq2} to first-order derivatives, defining new variables $\overline{\varphi}$ and $\overline{u}$ for the shear rates:

\eqn{\dfrac{\p^2 \varphi}{\p t^2} = \dfrac{\p \overline{\varphi}}{\p {y_{\lambda}}},
\qquad\qquad\qquad \overline{\varphi} = \dfrac{\p \varphi}{\p {y_{\lambda}}},
\label{FD-PDE-solid}}

\eqn{\dfrac{\p u}{\p t} = \dfrac{\p \overline{u}}{\p {y_{\delta}}},
\qquad\qquad\qquad \overline{u} = \dfrac{\p u}{\p {y_{\delta}}}.
\label{FD-PDE-fluid}}

Equations \eref{FD-PDE-solid} and \eref{FD-PDE-fluid} are discretised on a fixed grid along the wall-normal direction since all solid and fluid displacements are planar. The solid-liquid interface remains therefore planar during the motion and there is no need for numerical Lagrangian tracking of the interface. We use a backward difference approximation in time with an index $j\geq 0$ and step size $\Delta t$, and a centred difference approximation in space with an index $0\leq k \leq K$, and step sizes $\Delta y_{\lambda}$ and $\Delta y_{\delta}$ for the solid and the fluid, respectively. The top of the solid grid is $k=I_s$ and the bottom of the fluid grid is $k=I_f=I_s+1$, these grid positions representing the same point in physical space. For $j \geq 2$, second-order approximations are used for \eref{eq1} and \eref{eq2}. At the interface ($k=I_s$ and $k=I_f$), first-order spatial derivatives are discretised using a backward difference approximation in the solid and a forward difference approximation in the fluid. For the exterior boundaries of the grid, first-order approximations for the spatial derivatives are used to discretise \eref{bc1} and \eref{bc4}. In order to discretise \eref{bc2} and \eref{bc3}, both grid points $k=I_s$ and $k=I_f$ are used:

\eqn{u_{I_f}^j=\frac{\varphi_{I_s}^{j-2}-4 \varphi_{I_s}^{j-1}+3\varphi _{I_s}^j}{2\Delta t},\label{FD-velocity-BC} \quad \bar{u}_{I_f}^j=\rho L \bar{\varphi}_{I_s}^j.}

This discretisation is given explicitly in Appendix \ref{fd-appendix}. The system is arranged in a block tri-diagonal matrix, with values at the previous time steps, $j-1$ and $j-2$, forming the other side of a matrix equation. For $j=0$, the stationary initial conditions \eref{ic3} are imposed. For $j=1$, the discretisations are altered to use first-order approximations for all the first-order time derivatives. For the second-order time derivatives in \eref{FD-PDE-solid}, \eref{ic3} are incorporated via a ghost point at $j=-1$. In order to maintain the tri-diagonal form of the matrix, the discretisation of the spatial derivatives at the material interface is first-order.

\begin{figure}[h]
  \begin{center}
    \begin{tabular}{cc}
    \includegraphics[]{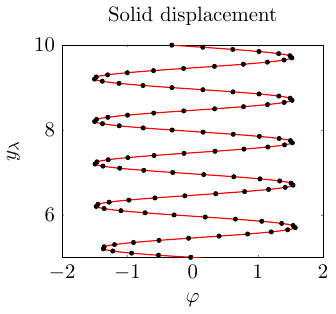} & \includegraphics[]{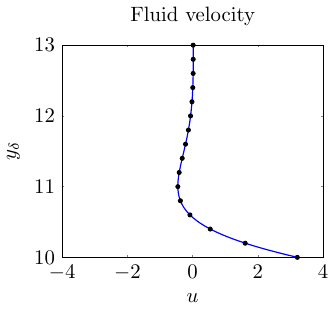}
    \end{tabular}
    \vspace{-7mm}
  \end{center}
  \caption{Comparison between the numerical solutions (circles) and the transient analytical solutions (lines).\label{numerical-checks}}
  \vspace{-7mm}
\end{figure}

Figure \ref{numerical-checks} shows the numerical solutions and the analytical solutions \eref{transient-solid-soln} and \eref{transient-fluid-soln}, at $t=15$, with $\tau>15$, $\rho L=1$, $h_{\lambda}=h_{\delta}=10$. For small enough numerical step sizes $\Delta t$, $\Delta y_\lambda$ and $\Delta y_\delta$, the numerical and analytical solutions overlap. Defining the error to be the absolute difference between the numerical results and the exact values computed from \eref{transient-solid-soln} and \eref{transient-fluid-soln}, the overall accuracy of the scheme has been found to be between order one and order two with respect to $\Delta t$, $\Delta y_{\lambda}$ and $\Delta y_{\delta}$, for a variety of nondimensional parameters.

%% file: include/5-physical-results.tex
\section{Physical results}
\label{physical-results}

The primary motivation for solving the system \eref{eq1} - \eref{ic3} analytically in \S\ref{analytical-results} is to gain insight into the separate contributions to the transient motion, which would not be possible by use of the numerical solution only. In \S\ref{physical-results-analytical} we discuss the physical results obtained from the analytical solutions \eref{transient-solid-soln} and \eref{transient-fluid-soln}. The solutions are investigated graphically in \S\ref{physical-results-graphical} and in the context of relevant technological applications in \S\ref{physical-results-parameters}. In \S\ref{physical-results-transient} we quantify the timescale for the evolution of the transient profiles into the periodic form found in \S\ref{section-fourier}.

\subsection{Physical interpretation of the analytical solutions}
\label{physical-results-analytical}

\subsubsection*{Classification of summation terms}

\begin{sloppypar}
    The solutions \eref{transient-solid-soln} and \eref{transient-fluid-soln} quantitatively describe a set of evolving elastic wave reflections in the solid layer for $0\leq y_{\lambda} \leq h_{\lambda}$ and a set of damped oscillations in the fluid layer for $y_{\delta}\geq h_{\delta}$. Successive Heaviside functions become non-zero at different times and locations, whereby more terms in the series contribute to the summations for $\varphi$ and $u$ and result in a superposition of oscillatory motions in the layers. For $0\leq t<\tau$, all of the Heaviside functions involving $\tau$ are zero, leaving three distinct types of terms contributing to the summations: those with Heaviside functions whose arguments contain $t-y_{\lambda}$ or $t+y_{\lambda}$ in \eref{transient-solid-soln} for the solid displacement $\varphi$, and those with Heaviside functions whose arguments depend only on $t$ in \eref{transient-fluid-soln} for the fluid velocity $u$, each multiplying a corresponding $G_{\mathcal{S},m,n}$ or $G_{\mathcal{F},m,n}$ function. The two types of terms in \eref{transient-solid-soln} involving $t-y_{\lambda}$ and $t+y_{\lambda}$ correspond respectively to wavefronts of displacement in the solid travelling upwards from $y_{\lambda}=0$ to $y_{\lambda}=h_{\lambda}$ and wavefronts travelling downwards from $y_{\lambda}=h_{\lambda}$ to $y_{\lambda}=0$. The terms in \eref{transient-fluid-soln} correspond to the shear motion transmitted into the fluid by the set of upward travelling waves in the solid when these elastic waves reach $y_{\lambda}=h_{\lambda}$. These terms depend on $y_{\delta}$ only via the $G_{\mathcal{F},m,n}$ functions as defined in \eref{GF0-exact}, rather than via the Heaviside functions. The three types of profile are depicted qualitatively in figure \ref{geometry}.

\end{sloppypar}

\subsubsection*{Evolution of reflections and transmissions}

\begin{sloppypar}

    The index $n$ denotes the number of partial reflections that occur at the material interface located at $y_{\lambda}=h_{\lambda}$. The first upward travelling term with $n=0$, that is, $\mathrm{H}\lb t-y_{\lambda}\rb G_{\mathcal{S},0,0}\lb t-y_{\lambda}\rb$, is equal to $\mathrm{H}\lb t-y_{\lambda}\rb\sin{\lb\omega t\rb}$. This term is the incident sinusoidal wave due to the imposed forcing at $y_{\lambda}=0$. When $t=h_{\lambda}$, this wave reaches the interface at $y_{\lambda}=h_{\lambda}$ for the first time. For $t > h_{\lambda}$, the first transmission term in \eref{transient-fluid-soln} for the fluid and the first downward travelling term in \eref{transient-solid-soln} for the solid both switch on. At $t=2h_{\lambda}$, the first downward travelling wave reaches $y_{\lambda}=0$ and the second upward travelling term, with $n=1$, switches on. For $t>2h_{\lambda}$, each subsequent upward travelling wave switches on at $y_{\lambda}=0$ when $t$ is an even multiple of $h_{\lambda}$ and reaches $y_{\lambda}=h_{\lambda}$ when $t$ is an odd multiple of $h_{\lambda}$. Each downward travelling wave switches on at $y_{\lambda}=h_{\lambda}$ when $t$ is an odd multiple of $h_{\lambda}$ and reaches $y_{\lambda}=0$ when $t$ is an even multiple of $h_{\lambda}$. Each transmission term in \eref{transient-fluid-soln} switches on in the fluid when $t$ is an odd multiple of $h_{\lambda}$ for $y_{\delta}\geq h_{\delta}$. As $t$ increases, there are more non-zero terms in the summation, and the amplitudes of each term generally decrease with increasing $n$, so that at large $t$ the incremental change given by each additional term is small. After several reflections, the full superposed solution converges towards a time-periodic state in both layers.

\end{sloppypar}

\subsubsection*{Transient properties of the elastic wavefronts and shear-driven fluid layers}

\begin{sloppypar}

    For the travelling elastic waves in the solid, the dependence on $y_{\lambda}$ is contained in the arguments of the Heaviside functions and in the time-shifted coordinates in the $G_{\mathcal{S},0,n}\lb t\rb$ functions. For a fixed $y_{\lambda}$, if the Heaviside function for a particular wave in the summation is non-zero, the time-dependent shear displacement is thus given in \eref{GS0-exact} by $G_{\mathcal{S},0,n}\lb t- y_{\lambda} -2n h_{\lambda}\rb$ or $G_{\mathcal{S},0,n}\lb t+ y_{\lambda}-2\lb n+1 \rb h_{\lambda}\rb$. The first two sinusoidal terms in \eref{GS0-exact} are periodic in time, but the integrand in the last term contains a factor that decays exponentially in time. There is thus a decaying transient contribution to each solid wave that travels with the advancing wave due to the Heaviside functions. An exception is the term with $n=0$ because the transient term is identically zero in that case. This incident wave is a simple translational wave that is solely periodic in time. The transmitted motion in the fluid, given by the linear superposition of the $G_{\mathcal{F},0,n}\lb y_{\delta},t \rb$ functions, exhibits a similar separation into periodic and transient terms. Noting that $y_{\delta}-h_{\delta}$ is the vertical coordinate measured relative to the interface at $y_{\delta}=h_{\delta}$, the first term in \eref{GF0-exact} consists of a weighted sum of upward travelling sinusoids, with an amplitude that decays exponentially with the distance from the interface because of viscous effects. This part of the solution is similar to the Stokes layer solution for a shear-driven fluid \cite{stokes-1851}. The second term in \eref{GF0-exact} is transient, containing a term that decays exponentially in time.

\end{sloppypar}

\subsubsection*{Transient behaviour at the end of the forcing}

\begin{sloppypar}
    Our discussion has been confined to $t<\tau$, with all of the terms involving $\tau$ in \eref{transient-solid-soln} and \eref{transient-fluid-soln} remaining zero until the forcing at $y_{\lambda}=0$ ends at $t=\tau$. These terms correspond to the time-dependent behaviour after the forcing stops. For $t>\tau$, they become non-zero and destructively interfere with the non-zero terms existing for $0<t<\tau$. The factors $\cos\lb\omega\tau\rb$ and $\sin\lb\omega\tau\rb$ determine which of these new terms contribute the most to the transient behaviour at the end of the forcing, with a special case occurring when the forcing is switched off after a whole number of periods (i.e. $\omega\tau$ is a multiple of $2\pi$). In this case, $\cos{\lb\omega\tau\rb}=1$ and $\sin{\lb\omega\tau\rb}=0$, and the additional terms for $t>\tau$ are identical to those for $t>0$, with a time shift $\tau$.
\end{sloppypar}

\subsection{Visualisation of the system dynamics}
\label{physical-results-graphical}

In this section we discuss plots of the analytical solutions \eref{transient-solid-soln}, \eref{transient-fluid-soln}, \eref{periodic-solid-soln} and \eref{periodic-fluid-soln}, in order to establish qualitative transient properties of the solutions over short and long timescales which cannot be obtained from the analytical solutions only. We first analyse the initial motion of both layers after the start of the forcing at $y_{\lambda}=0$, consisting of the incident wave, the first reflection in the solid and the first transmission into the fluid, as discussed in \S\ref{physical-results-analytical}. These three initial profiles share properties with the three types of profile arising in subsequent reflections. They are therefore representative of the system dynamics until $t=\tau$.

Figure \ref{plot-1-switch-on} shows the motion of both layers with $h_{\lambda}=10$, $\rho L=1$ and $t\ll\tau$. The fluid layer remains at rest until $t=10$, when the periodic and transient velocity contributions contained in \eref{GF0-exact} both become non-zero for $y_{\delta}\geq h_{\delta}$. As shown in figure \ref{plot-2-short-transient-fluid}, the two contributions cancel completely at $t=10$, the instant when the reflection occurs, but as the transient contribution starts to decay the overall velocity starts to grow near the interface whilst remaining zero further away. The periodic contribution quickly begins to dominate and the motion of the fluid layer becomes periodic with its amplitude decaying away from the interface. This cancellation, with a transient term receding to reveal a periodic profile, also occurs in the transient extension to the Stokes second problem \cite{panton-1968} and in similar heat transfer problems \cite{carslaw-jaeger-1959}. The fluid remains stationary far from the interface at all times, beyond a distance of around $y_\delta-h_\delta>3$, i.e. $\dimn{y}-\dimn{h}>3\dimn{\delta}$, consistent with the assumption of an unbounded fluid layer in \S\ref{mathematical-formulation}.

\input{include/plots/5b-1-switch-on}

\input{include/plots/5b-2-short-transient-fluid}

For $t>10$, after the reflection at the interface, the superposed elastic motions of the incident and reflected waves result in an overall solid displacement that is larger than that of the incident wave alone. As the first reflected wave travels downwards from $y_{\lambda}=h_{\lambda}$ to $y_{\lambda}=0$, for $t>10$, the periodic and transient displacement contributions contained in \eref{GS0-exact} both become non-zero. The transient contribution is largest at the front of the wave, whilst the area behind the advancing wave is dominated by the periodic contribution. Unlike the transient contributions to the fluid velocity, the solid transient contribution does not have an oscillatory shape. As shown in figure \ref{plot-2-short-transient-solid}, the transient contribution appears steady in a frame of reference moving with the reflected wavefront, so that its shape remains unchanged as it travels downwards.

\input{include/plots/5b-2-short-transient-solid}

For $t>20$, the superposition of further reflected elastic waves in the solid and transmissions into the fluid results in a second kind of transient evolution. This evolution can be visualised by the displacement and velocity at the interface, as shown in figure \ref{plot-3-wall-velocity}. The interface does not move until $t=10$, after which the velocity oscillates due to the forcing at $y_{\lambda}=0$. At $t=30$, when the second reflection occurs, the addition of another transmitted layer causes the velocity to obtain a larger amplitude. For the third reflection at $t=50$, the incremental change to the interface velocity is smaller, and, as more reflections occur, the amplitude approaches that of the solution \eref{periodic-fluid-soln-merged} found by assuming periodic motion. Furthermore, by comparing the periodic solutions \eref{periodic-solid-soln} and \eref{periodic-fluid-soln} to the transient solutions \eref{transient-solid-soln} and \eref{transient-fluid-soln}, figure \ref{plot-4-long-transient} shows that the whole two-layer system approaches a periodic state to which the summations in \eref{transient-solid-soln} and \eref{transient-fluid-soln} have converged. This convergence occurs after around eight reflections when $t=160$. The long transient evolution to the periodic state is investigated further in \S\ref{physical-results-transient}.

\input{include/plots/5b-3-wall-velocity}
\input{include/plots/5b-4-long-transient}

When $t\geq\tau$, the motion of the two layers is no longer driven by oscillation of the lower boundary at $y_{\lambda}=0$. Figure \ref{plot-5-switch-off} shows the dynamics of the same system ($h=10$, $\rho L=1$, $\tau=200$) after the oscillation at $y_{\lambda}=0$ stops. The motion of the two layers is periodic at the switch-off time $t=200$ and, for $200<t<500$, it evolves back to the initial state of the system, i.e. $\varphi=0$ and $u=0$.

\input{include/plots/5b-5-switch-off}

\subsection{Parameter dependence}
\label{physical-results-parameters}
 
The dependence of the displacement and velocity profiles on the physical parameters is studied for three representative cases related to applications of shear waves in ultrasound viscometry \cite{manfredi-etal-2019}, active methods for turbulent drag reduction \cite{bird-santer-morrison-2018} and SAW-based biosensors \cite{lange-etal-2008}. Although only sinusoidal forcing is studied herein, the linearity of the system implies that the results can be applied to more complex forcing patterns. In this section we only consider the periodic solutions \eref{periodic-solid-soln} and \eref{periodic-fluid-soln}, so that transient effects do not affect the parameter dependence. For a better interpretation of the results in view of the applications, the distance $\dimn{y}$ is scaled herein with the thickness of the solid $\dimn{h}$, while $\dimn{\varphi}$ and $\dimn{u}$ are scaled with the displacement and velocity amplitudes of the imposed forcing at $\dimn{y}=0$, i.e. $\dimn{\varphi}_{\text{wall}}=\dimn{\varphi}_{0}$ and $\dimn{u}_{\text{wall}}= 2\pi \dimn{\varphi}_{0} / \dimn{T}_0$.

\subsubsection*{Application 1: Viscometry}

\begin{figure}[h]
    \begin{center}
    \begin{tabular}{cc}
        \includegraphics[]{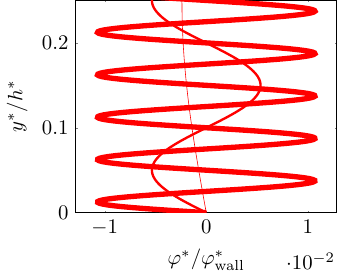} &
        \includegraphics[]{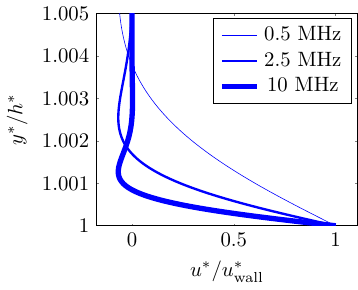}
    \end{tabular}
    \vspace{-7mm}
\end{center}
    \caption{Dependence of solid displacement (red) and fluid velocity (blue) on forcing frequency, for a 6.28mm aluminium block underneath a layer of 850 kg/m\textsuperscript{3} oil with viscosity 300 mPa s \cite{manfredi-etal-2019}.\label{c1-frequency-profiles}}
    \vspace{-5mm}
\end{figure}

Ultrasound viscometry setups use oils of varying densities and viscosities, and water for calibration \cite{manfredi-etal-2019}. Forcing frequencies span the 0.5-15 MHz range, using a thin aluminium layer underneath an oil layer. As shown in figure \ref{c1-frequency-profiles}, a higher frequency results in a shorter wavelength in the solid and in a larger displacement amplitude. In the fluid, higher forcing frequencies result in a smaller penetration depth, i.e. a fluid motion that is closer to the solid-liquid interface.

\begin{figure}[h]
    \begin{center}
    \begin{tabular}{c}
        \includegraphics[]{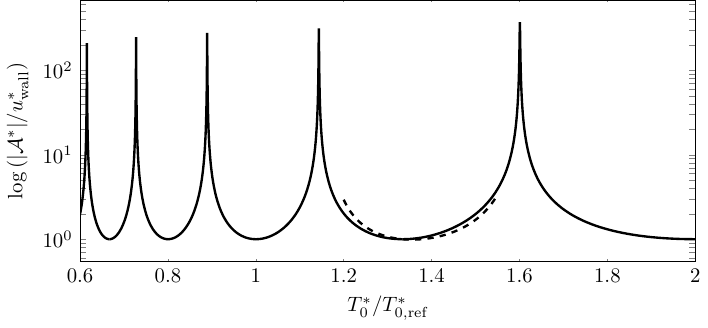}
    \end{tabular}
    \vspace{-7mm}
    \end{center}
    \caption{Dependence of interface velocity amplitude on $\dimn{T_0}$ (solid), compared to $T^*_{0,\text{ref}}=1$ MHz with a series expansion around one of the minima (dashed).\label{c1-frequency-amplitude}}
    \vspace{-5mm}
\end{figure}

For each of the chosen frequencies in figure \ref{c1-frequency-profiles}, the maximum velocity at the interface happens to be close to the forcing amplitude $\dimn{u}_{\text{wall}}$. The maximum wall velocity $\dimn{\mathcal{A}}$ can however be much larger than $\dimn{u}_{\text{wall}}$, with finite maximum values occurring around critical resonant frequencies, as shown in figure \ref{c1-frequency-amplitude}. This result demonstrates that the solid-fluid system behaves as a forced resonator that is damped by the fluid viscous effects.
The critical resonant frequencies, occurring when $\omega h_{\lambda}$ is an integer multiple of $\pi$, are also the frequencies at which $\dimn{\mathcal{A}}$ changes sign, although $|\dimn{\mathcal{A}}|$ is plotted for clarity. Using the analytical forms of $\mathcal{A}$ and $\mathcal{T}$ in \eref{periodic-fluid-soln-merged-amplitude}, $\dimn{\mathcal{A}}$ may be written as:

\eqn{
|\mathcal{A}^*|= \dfrac{\varphi_0^*}{T_0^*} \omega^{3/2} 
\dfrac{\rho L |\sin{\lb\omega h_{\lambda}\rb|}
\sqrt{\lb\dfrac{\rho L}{\sqrt{\pi}}
\cot{\lb\omega h_{\lambda}\rb}
-1\rb^2+1}}{\lb \rho L \cos{\lb\omega h_{\lambda}\rb} - \sqrt{\pi} \sin{\lb\omega h_{\lambda}\rb} \rb^2 + \pi \lb \sin{\lb\omega h_{\lambda}\rb} \rb^2}.
\label{aDimn}
}

Noting that both $h_{\lambda}$ and $L$ depend implicitly on $\dimn{T_0}$, we may study the behaviour of $\dimn{\mathcal{A}}$ asymptotically in the proximity of the minima. Rewriting $\sin{\lb\omega h_{\lambda}\rb}$ and $\cos{\lb\omega h_{\lambda}\rb}$ as Taylor series around $\omega h_{\lambda}=k \pi$ where $k\in\mathbb{Z}^+$ (i.e. where $\sin{\lb\omega h_{\lambda}\rb}\to 0$), this expansion leads to a valid approximation of $\dimn{\mathcal{A}}\lb\dimn{T_0}\rb$ around the minima $\dimn{T_0}=2 \dimn{h}\sqrt{\dimn{\rho_s}/\dimn{\mu_s}}/k$. This expansion is shown as a dashed line in figure \ref{c1-frequency-amplitude} for one of the minima.

\begin{figure}[h]
    \begin{center}
    \begin{tabular}{c}
        \includegraphics[]{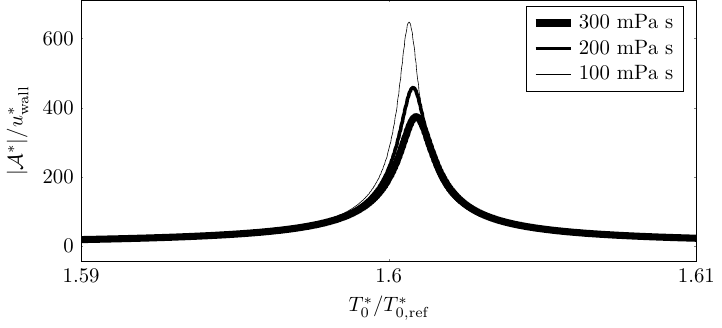}
    \end{tabular}
    \vspace{-7mm}
    \end{center}
    \caption{Effect of reducing $\dimn{\nu}$ on damped resonances around critical forcing frequencies.\label{c1-frequency-res}}
    \vspace{-5mm}
\end{figure}

We now consider the effect of varying $\dimn{\nu}$ on the maxima of $\dimn{\mathcal{A}}$, since the dependence of $\dimn{\mathcal{A}}$ on $\dimn{T_0}$ takes the form of damped resonance. The dynamic viscosity value of 300 mPas used in figure \ref{c1-frequency-amplitude} is reduced and kept within the range expected for test oils used in ultrasound viscometry applications. As shown in figure \ref{c1-frequency-res}, for lower viscosity (less damping) the maxima are larger and occur at a lower value of $\dimn{T_0}$. The limiting case of undamped resonance may be studied by considering the behaviour of $\dimn{\mathcal{A}}$ as $\dimn{\nu}\to 0$, or equivalently $L\to \infty$ in \eref{aDimn}. Analytically, $\lim_{L\to\infty} |\dimn{\mathcal{A}}|$ is proportional to $\sec{\lb\omega h_{\lambda}\rb}$, confirming that for an undamped system the maxima are resonant singularities occurring at $\omega h_{\lambda}=\left(2k-1\right) \pi /2$, or equivalently $\dimn{T_0}=4 \dimn{h}\sqrt{\dimn{\rho_s}/\dimn{\mu_s}} / \left(2k-1\right)$.

\subsubsection*{Application 2: Drag reduction}

\begin{figure}[h]
    \begin{center}
    \begin{tabular}{cc}
        \includegraphics[]{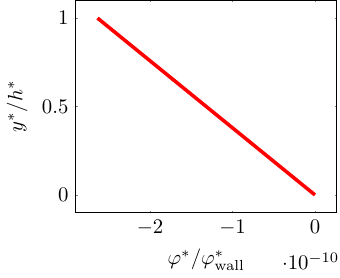}&
        \includegraphics[]{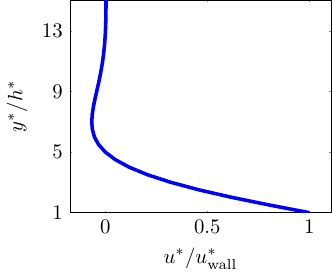}
    \end{tabular}
    \vspace{-7mm}
    \end{center}
    \caption{Solid displacement (red) and fluid velocity (blue) for a typical drag reduction laboratory setup, with 0.35mm layer of silicone, 6 Hz forcing and an adjacent layer of air \cite{bird-santer-morrison-2018}.\label{c2-thickness-profiles}}
\vspace{-5mm}
\end{figure}

\begin{figure}[h]
    \begin{center}
    \begin{tabular}{c}
        \includegraphics[]{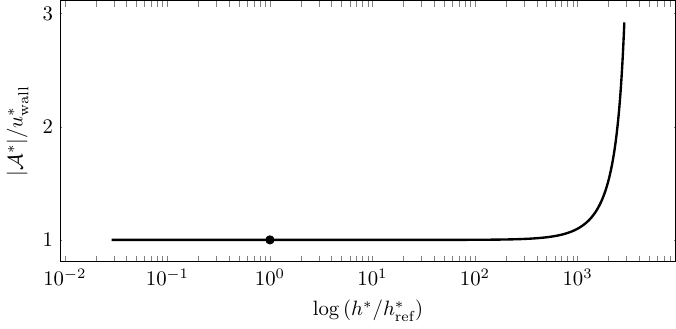}
    \end{tabular}
    \vspace{-7mm}
    \end{center}
    \caption{Dependence of maximum interface velocity on solid thickness, compared to $h^*_{\text{ref}}=0.35$ mm for a typical drag reduction setup \cite{bird-santer-morrison-2018} (black dot).\label{c2-thickness-amplitude}}
    \vspace{-5mm}
\end{figure}

For the travelling-wave method of turbulent drag reduction studied in \cite{bird-santer-morrison-2018}, a very thin elastic layer is stretched over a deformable lattice. The wall turbulence flows over the elastic surface. The low forcing frequencies used in such laboratory setups and the small thickness of the solid layer imply that the transverse elastic wavelength greatly exceeds the latter. This scenario approaches the behaviour of the fluid-only simplification adopted in \S\ref{section-fluid-only}. As shown in figure \ref{c2-thickness-profiles}, the solid displacement exhibits a very small linear shear throughout the layer rather than varying sinusoidally, the interface velocity is the same as that at the wall, and the penetration depth into the fluid is deeper than the solid layer itself. Controlling the fluid penetration depth is important in active drag reduction methods in order to ensure optimal interaction between the imposed shear waves and the turbulent structures. By extending the solid thickness well beyond the experimental values, the velocity amplitude at the interface grows and eventually approaches the resonance condition, as shown in figure \ref{c2-thickness-amplitude}.

\subsubsection*{Application 3: Biosensors}

\begin{figure}[h]
    \begin{center}
    \begin{tabular}{cc}
        \includegraphics[]{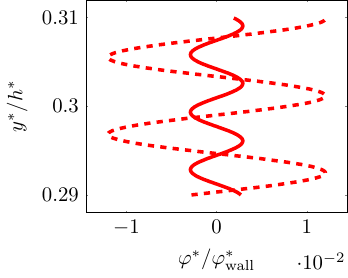}&
        \includegraphics[]{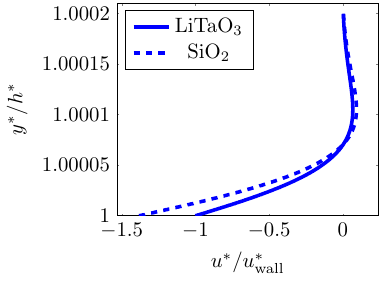}
    \end{tabular}
    \vspace{-7mm}
    \end{center}
    \caption{Solid displacement (red) and fluid velocity (blue) for a two materials used in biosensing applications \cite{lange-etal-2008}, with a 400 micron solid layer, 1 GHz forcing and an adjacent layer of water.
    \label{c3-bio-profiles}}
\vspace{-5mm}
\end{figure}

SAW-based devices used for biological sensing use shear waves in a solid substrate, with relevant biological material suspended in an adjacent fluid layer \cite{lange-etal-2008}. A variety of substrates are utilised, including lithium tantalate (LiTaO\textsubscript{3}) and quartz (SiO\textsubscript{2}), with thicknesses of the order of a few hundred microns. Forcing frequencies are even higher than in the viscometry case, ranging from 100~MHz to 3~GHz. Due to these high frequencies, resonance conditions are very closely spaced and small variations in solid density or shear modulus result in a significant change in the maximum interface velocity $\dimn{\mathcal{A}}$. The penetration depth into the fluid is very small, as shown in figure \ref{c3-bio-profiles}.

\subsection{Duration of long-term transients}
\label{physical-results-transient}

Due to the complexity of the fully transient solution, it is not always practical to extract information about the long-time transient dynamics from graphical analysis. The timescale on which the transient solutions \eref{transient-solid-soln} and \eref{transient-fluid-soln} approach the periodic solutions \eref{periodic-solid-soln} and \eref{periodic-fluid-soln} can be quantified numerically by considering the power balance of the fluid layer. Multiplying \eref{eq2} by the fluid velocity and integrating along the wall-normal direction, the balance involves the kinetic energy of the fluid, the energy input from the solid layer, and the energy dissipated due to viscous effects:

\eqn{\underbrace{\dfrac{1}{2}\dfrac{\mathrm{d}}{\mathrm{d}t}\int_{h_{\delta}}^{\infty} u^2\mathrm{d}y_{\delta}}_{\text{time rate of change of kinetic energy}} = \underbrace{{\left. -u\dfrac{\p u}{\p y_{\delta}}\right|}_{y_{\delta}=h_{\delta}}}_{\text{power input}}-\underbrace{\int_{h_{\delta}}^{\infty} \lb \dfrac{\p u}{\p y_{\delta}}\rb^2 \mathrm{d}y_{\delta}}_{\text{power dissipation}}.}

\begin{figure}[h]
    
    \begin{center}
        \begin{tabular}{cc}
        \includegraphics{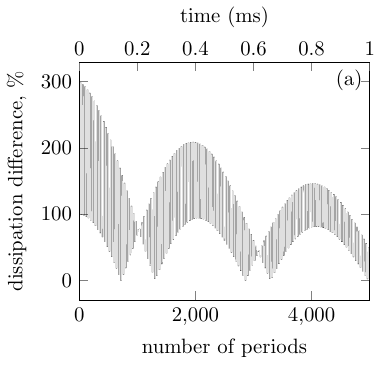}&
        \includegraphics{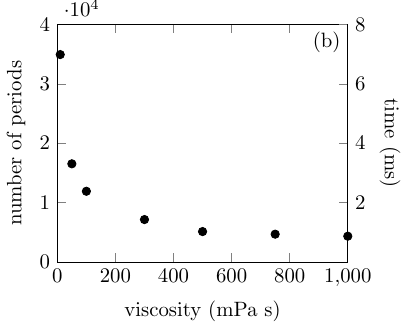}
        \end{tabular}
    \end{center}
    \caption{(a): Percentage difference between periodic and transient dissipation, averaged over each period. (b): Number of periods to reach $\leq 1\%$ agreement between the periodic and the transient solutions, for varying viscosity. Both plots are based on parameter values from the viscometry case studied in \S5(c), using a forcing frequency of 5 MHz.
    \label{5d-energy-plot}}
\vspace{-5mm}
\end{figure}

We monitor the viscous dissipation in the fluid to compute the overall transient evolution and compare the transient and periodic solutions. This choice is dictated by the dissipation in the fluid being always positive and relevant to the viscometry case studied in \S\ref{physical-results-parameters}, and by the absence of dissipation in the elastic layer. The total energy dissipated in the fluid in each period using the periodic solution \eref{periodic-fluid-soln} is constant, whereas the dissipation in each period using the transient solution \eref{transient-fluid-soln} varies as the series develops. Figure \ref{5d-energy-plot}(a) shows the difference between the viscous dissipation in periodic conditions and the transient viscous dissipation averaged over each period for typical values of the viscometry case.

For a range of viscosity values, forced at 5 MHz with a 6.28mm aluminium layer, the number of periods required for the maxima of the dissipation difference to decrease below $1\%$ varies from 4000 to 35000, with more viscous fluids (therefore more heavily damped systems) reaching a periodic state quicker, as shown in figure \ref{5d-energy-plot}(b).

For comparison, the drag-reduction case studied in figure \ref{c2-thickness-profiles} requires 169 periods for a 6~Hz forcing, while the lithium tantalate biosensor studied in figure \ref{c3-bio-profiles} requires 251846 periods for a 1 GHz forcing. Typical durations of transients are of the order of $10^{-3}$~s for ultrasound viscometry, $10$~s for travelling-wave drag reduction methods and $10^{-4}$ s in biosensors. Quantifying the duration of transient effects will aid in the design of future experiments and will establish whether periodic motion may be assumed.

%% file: include/plots/5b-1-switch-on.tex
\begin{figure}[h]
    \begin{center}
    \begin{tabular}{ccc}
    
    \includegraphics[]{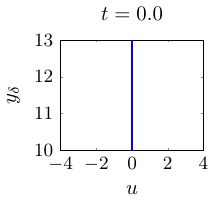} &
    \includegraphics[]{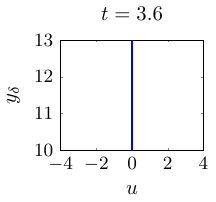} &
    \includegraphics[]{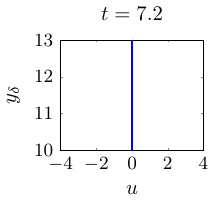}
    \vspace{-3mm}
    \\
    
    \includegraphics[]{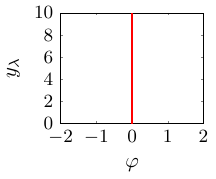} &
    \includegraphics[]{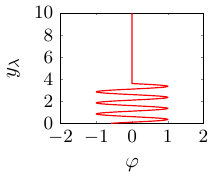} &
    \includegraphics[]{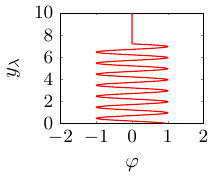}
    \vspace{-3mm}
    \\

    \includegraphics[]{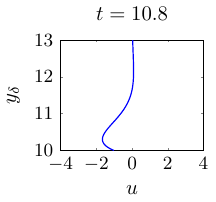} &
    \includegraphics[]{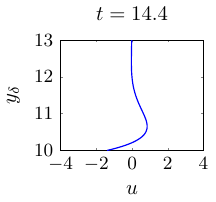} &
    \includegraphics[]{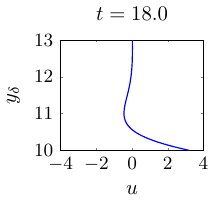}
    \vspace{-3mm}
    \\
    
    \includegraphics[]{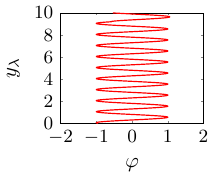} &
    \includegraphics[]{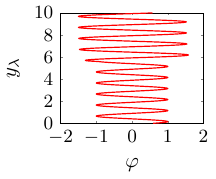} &
    \includegraphics[]{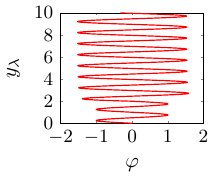}
    \vspace{-3mm}
    \\

    \end{tabular}
    \vspace{-5mm}
    \end{center}
    \caption{Plots of solid displacement (red) and fluid velocity (blue) for the first transmission and reflection.
    \label{plot-1-switch-on}}
    \vspace{-5mm}
    \end{figure}

%% file: include/plots/5b-2-short-transient-fluid.tex
\begin{figure}[h]
    \begin{center}
    \begin{tabular}{cc}
    
    \includegraphics[]{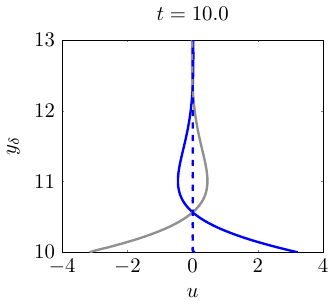} &
    \includegraphics[]{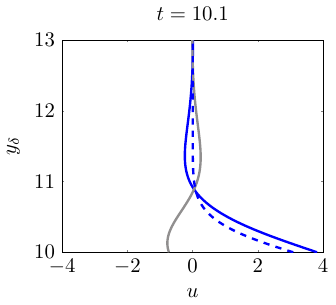} 
    \vspace{-2mm}
    \\
    
    \includegraphics[]{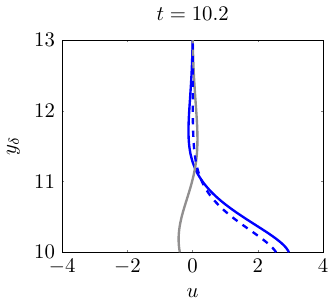} &
    \includegraphics[]{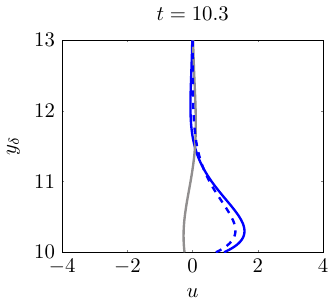} 
    
    \end{tabular}
    \vspace{-5mm}
    \end{center}
    \caption{Comparison of the periodic (solid blue) and transient (solid grey) contributions to the total fluid velocity (dashed blue), at the start of the first fluid transmission.
    \label{plot-2-short-transient-fluid}}
    \vspace{-5mm}
    \end{figure}

%% file: include/plots/5b-2-short-transient-solid.tex
\begin{figure}[h]
    \begin{center}
    \begin{tabular}{ccc}
    
    \includegraphics[]{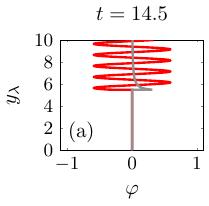} &
    \includegraphics[]{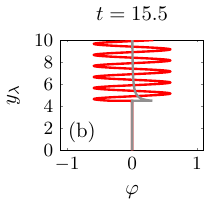} &
    \includegraphics[]{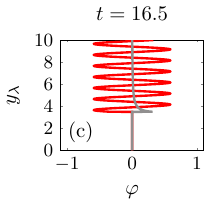}

    \end{tabular}
    \vspace{-2mm}
    \begin{tabular}{cccc}
    \qquad&
    \includegraphics[]{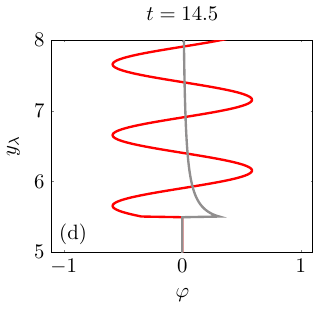} &
    \includegraphics[]{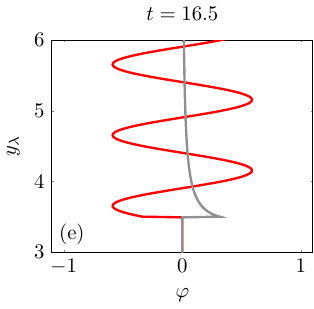}&
    \qquad
    
    \end{tabular}
    \vspace{-5mm}
    \end{center}
    \caption{Periodic (red) and transient (grey) contributions to the total solid displacement for the first reflected wave (without the superposed incident wave). (d) and (e) show the consistent profile shape in the vicinity of the wavefront.
    \label{plot-2-short-transient-solid}}
    \vspace{-5mm}
    \end{figure}

%% file: include/plots/5b-3-wall-velocity.tex
\begin{figure}[h]
    \begin{center}
    \begin{tabular}{cc}
    \includegraphics[]{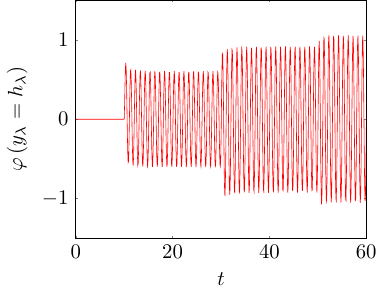} &
    \includegraphics[]{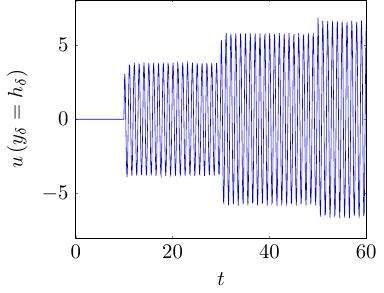}
    \end{tabular}
    \vspace{-5mm}
    \end{center}
    \caption{Development of the displacement and velocity at the solid-fluid interface.
    \label{plot-3-wall-velocity}}
    \vspace{-5mm}
    \end{figure}

%% file: include/plots/5b-4-long-transient.tex
\begin{figure}[h]
    \begin{center}
    \begin{tabular}{ccc}
    
    \includegraphics[]{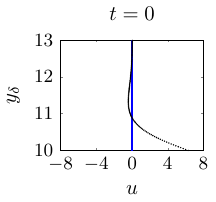} &
    \includegraphics[]{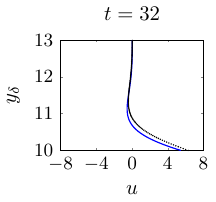} &
    \includegraphics[]{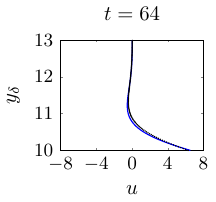}
    \vspace{-2mm}
    \\
    
    \includegraphics[]{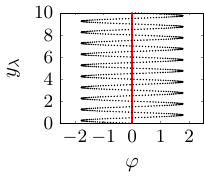} &
    \includegraphics[]{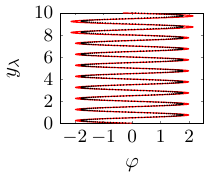} &
    \includegraphics[]{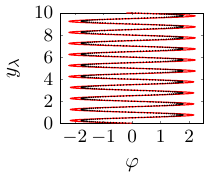}
    \vspace{-2mm}
    \\

    \includegraphics[]{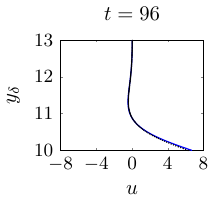} &
    \includegraphics[]{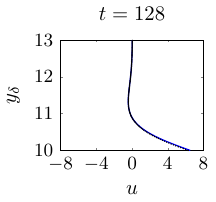} &
    \includegraphics[]{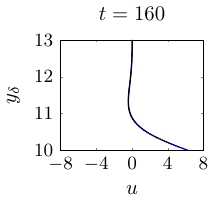}
    \vspace{-2mm}
    \\
    
    \includegraphics[]{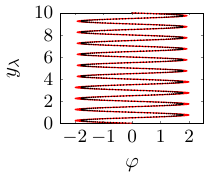} &
    \includegraphics[]{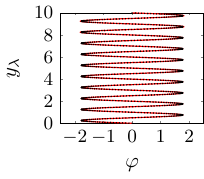} &
    \includegraphics[]{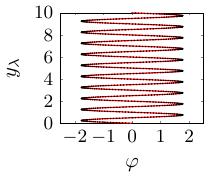}
    \vspace{-2mm}
    \\

    \end{tabular}
    \vspace{-5mm}
    \end{center}
    \caption{Development of the transient solid displacement (red) and fluid velocity (blue) towards periodic profiles (black).
    \label{plot-4-long-transient}}
    \vspace{-5mm}
    \end{figure}

%% file: include/plots/5b-5-switch-off.tex
\begin{figure}[h]
    \begin{center}
    \begin{tabular}{ccc}
    
    \includegraphics[]{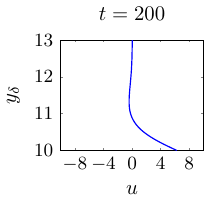} &
    \includegraphics[]{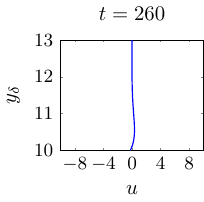} &
    \includegraphics[]{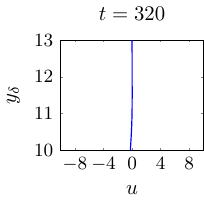}
    \vspace{-2mm}
    
    \\
    
    \includegraphics[]{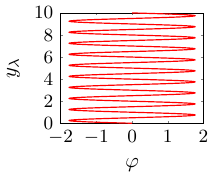} &
    \includegraphics[]{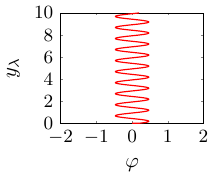} &
    \includegraphics[]{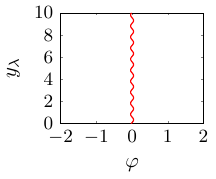}
    \vspace{-2mm}
    \\

    \includegraphics[]{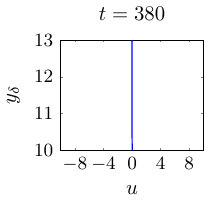} &
    \includegraphics[]{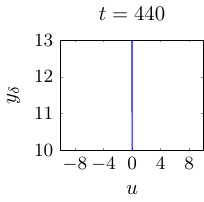} &
    \includegraphics[]{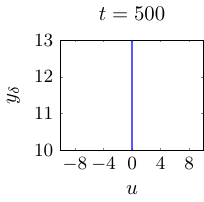}
    \vspace{-2mm}
    \\
    
    \includegraphics[]{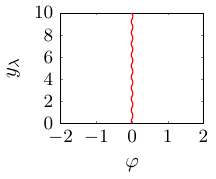} &
    \includegraphics[]{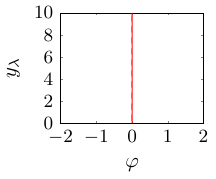} &
    \includegraphics[]{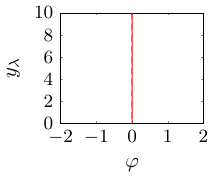}
    \vspace{-2mm}
    \\

    \end{tabular}
    \vspace{-5mm}
    \end{center}
    \caption{Return of the transient solid displacement (red) and fluid velocity (blue) towards the initial conditions, after the periodic forcing is switched off.
    \label{plot-5-switch-off}}
    \vspace{-5mm}
    \end{figure}

%% file: include/6-conclusions.tex
\section{Conclusions}
\label{conclusions}

In this paper, we have studied the dynamics of a coupled system consisting of a Newtonian fluid located on an elastic solid that is forced sinusoidally. The problem has been solved analytically and numerically. We have focused on the transient evolution from the beginning of the forced oscillations, solved by Laplace transforms, and on the periodic behaviour that ensues once the transient has vanished, solved by Fourier modes, presented in \S\ref{section-laplace} and \S\ref{section-fourier}. In \S\ref{section-fluid-only} we have demonstrated that these solutions reduce to the classic fluid mechanics results of the transient and periodic Stokes second problems in the case of a vanishingly thin solid layer. The analytical transient solution is revelatory of the dynamics because it is expressed as series summations that elucidate the propagation and reflections of the elastic waves and the viscous dissipation of the oscillatory motions in the viscous fluid. Integral terms pertain to the non-periodic behaviour in the solid and in the fluid. The periodic solution highlights that the system can be viewed as a resonant oscillator that is damped by the fluid viscous effects. The forcing periods at which resonance occurs are expressed in analytical form in the limit of vanishing fluid viscosity. The long-term duration of the transient effects has been quantified by considering the power balance of the fluid layer. These physical results, derived from the exact solutions, have been presented in \S\ref{physical-results}.

The interaction between shear-driven solids and viscous fluids is of interest due to their uses in engineering applications, including in-situ viscometry, turbulent drag reduction and manipulation of biological flows. We hope that our results will be useful as a theoretical framework to aid the design of future experiments. The two-layer linear system studied herein can serve as a limiting case for more complex systems, involving more realistic geometries featuring two- or three-dimensional effects, non-Newtonian liquids, bounded fluid layers and multiple solid layers. The problem can also be extended to non-sinusoidal forcing. Future research aims could be to quantify the duration and magnitude of transient effects in systems with pulsed forcing and to utilise the resonance effects to amplify the velocity or displacement at the solid/fluid interface.

%% file: include/A-contour-integration.tex
\section{Contour integration for transient solution}
\label{contour-integration}

\begin{figure}[h]
  \begin{center}
    \includegraphics[width=.5\textwidth]{"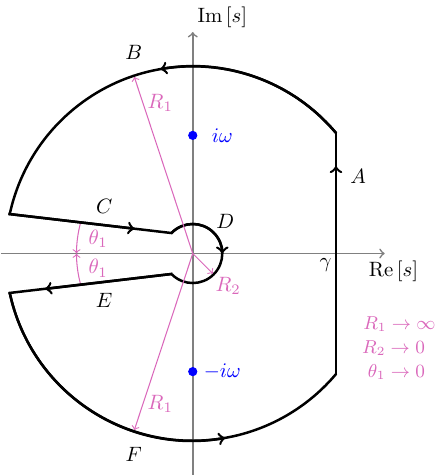"}
    \vspace{-3mm}
    \caption{Modified Bromwich contour for evaluating $G_{\mathcal{S},m,n}$ and $G_{\mathcal{F},m,n}$.}
    \label{contour}
    \vspace{-5mm}
  \end{center}
\end{figure}

In this appendix we present detailed derivations of the functions $G_{\mathcal{S},0,n}$, $G_{\mathcal{S},1,n}$, $G_{\mathcal{F},0,n}$ and $G_{\mathcal{F},1,n}$, found by integration in the complex plane along the contour $ABCDEF$ shown in figure \ref{contour}. By applying the definition of the inverse Laplace transform in order to invert \eref{GSmn-definition} and \eref{GFmn-definition}, integration must be performed along the Bromwich contour $s=\gamma\pm i \infty$, where $\gamma$ is larger than the real part of any singularities of the integrands:

\eqn{
  G_{\mathcal{S},m,n}\lb t\rb=\dfrac{1}{2 \pi \imag} \int_{\gamma-\imag\infty}^{\gamma+\imag\infty} \dfrac{ s^m \omega^{1-m}}{ s^2+\omega^2}\lb \dfrac{\rho L - \sqrt{s}}{\rho L + \sqrt{s}}\rb ^n e^{st} \mathrm{d}s,
  \label{GSmn-integral}
}
\eqn{
  G_{\mathcal{F},m,n}\lb y_{\delta}, t\rb= \dfrac{1}{2 \pi \imag} \int_{\gamma-\imag\infty}^{\gamma+\imag\infty} \dfrac{2 \rho L  s^{1+m} \omega^{1-m}  e^{\sqrt{s}\lb h_{\delta}-y_{\delta}\rb}}{\lb  s^2+\omega^2\rb \lb \rho L + \sqrt{s}\rb} \lb \dfrac{\rho L - \sqrt{s}}{\rho L + \sqrt{s}}\rb ^n e^{st} \mathrm{d}s.
  \label{GFmn-integral}
}

Calling $A$ the Bromwich contour for the integrals in \eref{GSmn-integral} and \eref{GFmn-integral}, the integration may be performed by closing the contour to the left and by applying the residue theorem \cite{brown-churchill-2009}. A branch cut is taken around the negative real axis and around the origin, so that $\sqrt{s}$ is single-valued along the whole contour \cite{boas-1999}. In the limiting case where the radius $R_1$ of the outer circular arcs approaches $\infty$, the radius $R_2$ of the circle around the origin approaches $0$, and the complex argument of the branch cut approaches $\pm \pi$ i.e. $\theta_1$ approaches $0$, the Bromwich contour is recovered. The integrands for both $G_{\mathcal{S},m,n}$ and $G_{\mathcal{F},m,n}$ contain poles at $s=\pm \imag \omega$. By the residue theorem, the integral along the contour $ABCDEF$ is equal to the sum of the residues of all poles. Allowing $\mathrm{I_A}$ to denote the integral along contour $A$ etc., and $\mathrm{Res}_{\lb s_0 \rb}$ to denote the residue at $s=s_0$, the residue theorem may be rearranged to find $\mathrm{I_A}$:

\eq{\mathrm{I_A}+\mathrm{I_B}+\mathrm{I_C}+\mathrm{I_D}+\mathrm{I_E}+\mathrm{I_F} = 2 \pi \imag \lb\mathrm{Res}_{\lb \imag\omega \rb}+\mathrm{Res}_{\lb -\imag\omega \rb}\rb}
\eq{\therefore \dfrac{\mathrm{I_A}}{2 \pi \imag} = \mathrm{Res}_{\lb \imag\omega \rb}+\mathrm{Res}_{\lb -\imag\omega \rb} -\dfrac{1}{2 \pi i} \lb \mathrm{I_B}+\mathrm{I_C}+\mathrm{I_D}+\mathrm{I_E}+\mathrm{I_F} \rb.}
The contributions to $\mathrm{I_A}$ from $\mathrm{I_B}$ and $\mathrm{I_F}$ tend to zero by Jordan's lemma as $R_1 \to \infty$, for both $G_{\mathcal{S},m,n}$ and $G_{\mathcal{F},m,n}$. The residues and the contributions from $\mathrm{I_C}$, $\mathrm{I_D}$ and $\mathrm{I_E}$ must be calculated explicitly for each integrand.

\subsection*{Contour integrals for $G_{\mathcal{S},m,n}$}

On $C$, $s=Re^{\imag \theta}$ as $\theta \to \pi$. Hence, $s=-R$, $\sqrt{s}=\imag\sqrt{R}$ and $\mathrm{d}s=-\mathrm{d}R$:
\eq{\mathrm{I_C} = \int_\infty^0 \dfrac{{\lb-R\rb}^m \omega^{1-m}}{R^2+\omega^2}\lb \dfrac{\rho L-\imag\sqrt{R}}{\rho L+\imag\sqrt{R}} \rb^n e^{-Rt} \lb -\mathrm{d}R \rb.}
On $E$, $s=Re^{\imag \theta}$ as $\theta \to -\pi$. Hence, $s=-R$, $\sqrt{s}=-\imag\sqrt{R}$ and $\mathrm{d}s=-\mathrm{d}R$:
\eq{\mathrm{I_E} = \int_0^\infty \dfrac{{\lb-R\rb}^m \omega^{1-m}}{R^2+\omega^2}\lb \dfrac{\rho L+\imag\sqrt{R}}{\rho L-\imag\sqrt{R}} \rb^n e^{-Rt} \lb -\mathrm{d}R \rb,}
\eq{\therefore \mathrm{I_C}+\mathrm{I_E}=\int_0^\infty \dfrac{{\lb-R\rb}^m \omega^{1-m}}{R^2+\omega^2}\lb \lb \dfrac{\rho L-\imag\sqrt{R}}{\rho L+\imag\sqrt{R}} \rb^n -\lb \dfrac{\rho L+\imag\sqrt{R}}{\rho L-\imag\sqrt{R}} \rb^n \rb e^{-Rt} \mathrm{d}R.}
On $D$, $s=R_2 e^{\imag \theta}$. Hence $\sqrt{s}=e^{\imag\theta /2}\sqrt{R_2}$ and $\mathrm{d}s=\imag R_2 e^{\imag \theta}\mathrm{d}\theta$:
\eq{\mathrm{I_D} = \int_\pi^{-\pi} \dfrac{R_2^m e^{m \imag\theta} \omega^{1-m}}{\omega^2+R_2^2 e^{2\imag\theta}} \lb \dfrac{\rho L-e^{\imag\theta /2}\sqrt{R_2}}{\rho L+e^{\imag\theta /2}\sqrt{R_2}}\rb^n e^{R_2 e^{\imag\theta}t} \imag R_2 e^{\imag\theta} \mathrm{d}\theta.}

When constructing the solution $G_{\mathcal{S},m,n}$, either $m=0$ or $m=1$. Hence, $\mathrm{I_D}\to 0$ as $R_2\to 0$. We compute the residues at the two poles $s=\pm \imag\omega$:
\eq{\mathrm{Res}_{\lb \imag\omega \rb} = {\lb \dfrac{s^m \omega^{1-m}}{s+\imag\omega} \lb \dfrac{\rho L-\sqrt{s}}{\rho L+\sqrt{s}} \rb^n e^{st}\rb}_{s\to \imag\omega}=  \dfrac{\imag^{m-1}}{2} \lb \dfrac{\rho L-\sqrt{\imag\omega}}{\rho L+\sqrt{\imag\omega}} \rb^n e^{\imag\omega t},}
\eq{\mathrm{Res}_{\lb -\imag\omega \rb} = {\lb \dfrac{s^m \omega^{1-m}}{s-\imag\omega} \lb \dfrac{\rho L-\sqrt{s}}{\rho L+\sqrt{s}} \rb^n e^{st}\rb}_{s\to -\imag\omega}=\dfrac{\lb -\imag\rb^{m-1}}{2} \lb \dfrac{\rho L-\sqrt{-\imag\omega}}{\rho L+\sqrt{-\imag\omega}} \rb^n e^{-\imag\omega t}.}

The solutions are:
\eq{
  \begin{split}
    G_{\mathcal{S},0,n}=&\dfrac{1}{2\imag}{\lb\dfrac{\rho L -\sqrt{\imag\omega}}{\rho L +\sqrt{\imag\omega}}\rb}^n e^{\imag\omega t}-\dfrac{1}{2\imag}{\lb\dfrac{\rho L -\sqrt{-\imag\omega}}{\rho L +\sqrt{-\imag\omega}}\rb}^n e^{-\imag\omega t}\\
    -&\dfrac{1}{2\pi \imag}\int_0^\infty \dfrac{\omega}{R^2+\omega^2}\lb\lb\dfrac{\rho L-\imag\sqrt{R}}{\rho L+\imag\sqrt{R}}\rb^n-\lb\dfrac{\rho L+\imag\sqrt{R}}{\rho L-\imag\sqrt{R}}\rb^n\rb e^{-Rt} \mathrm{d}R,
  \end{split}
  \label{contour-output-GS0}
}
\eq{
  \begin{split}
    G_{\mathcal{S},1,n}=&\dfrac{1}{2}{\lb\dfrac{\rho L -\sqrt{\imag\omega}}{\rho L +\sqrt{\imag\omega}}\rb}^n e^{\imag\omega t}+\dfrac{1}{2}{\lb\dfrac{\rho L -\sqrt{-\imag\omega}}{\rho L +\sqrt{-\imag\omega}}\rb}^n e^{-\imag\omega t}\\
    -&\dfrac{1}{2\pi \imag}\int_0^\infty \dfrac{-R}{R^2+\omega^2}\lb\lb\dfrac{\rho L-\imag\sqrt{R}}{\rho L+\imag\sqrt{R}}\rb^n-\lb\dfrac{\rho L+\imag\sqrt{R}}{\rho L-\imag\sqrt{R}}\rb^n\rb e^{-Rt} \mathrm{d}R.
  \end{split}
  \label{contour-output-GS1}
}

As complex conjugates appear, the solutions are written as:
\eq{
  G_{\mathcal{S},0,n}=2\mathrm{Re}\lbs-\dfrac{\imag}{2}{\lb\dfrac{\rho L -\sqrt{\imag\omega}}{\rho L +\sqrt{\imag\omega}}\rb}^n e^{\imag\omega t}\rbs-\dfrac{1}{2\pi \imag}\int_0^\infty 2\imag \mathrm{Im}\lbs\dfrac{\omega}{R^2+\omega^2}\lb\dfrac{\rho L-\imag\sqrt{R}}{\rho L+\imag\sqrt{R}}\rb^n \rbs e^{-Rt} \mathrm{d}R,
}
\eq{
  G_{\mathcal{S},1,n}=2\mathrm{Re}\lbs\dfrac{1}{2}{\lb\dfrac{\rho L -\sqrt{\imag\omega}}{\rho L +\sqrt{\imag\omega}}\rb}^n e^{\imag\omega t}\rbs+\dfrac{1}{2\pi \imag}\int_0^\infty 2\imag \mathrm{Im}\lbs\dfrac{R}{R^2+\omega^2}\lb\dfrac{\rho L-\imag\sqrt{R}}{\rho L+\imag\sqrt{R}}\rb^n \rbs e^{-Rt} \mathrm{d}R.
}
We rewrite the denominators of the complex fractions to isolate the overall real and imaginary components:
\eq{
\begin{split}  
G_{\mathcal{S},0,n}=
\mathrm{Re}\lbs \lb\dfrac{\rho^2 L^2-\omega-2\imag\rho L\sqrt{\omega /2}}{\lb\rho L+\sqrt{\omega /2}\rb^2+\omega/2}\rb^n \lb-\imag e^{\imag\omega t}\rb\rbs
\\-\dfrac{1}{\pi}\int_0^\infty\mathrm{Im}\lbs\dfrac{\omega\lb\rho L-\imag\sqrt{R}\rb^{2n}}{\lb R^2+\omega^2\rb\lb\rho^2 L^2+R\rb^n}\rbs e^{-Rt} \mathrm{d}R,
\end{split}
}
\eq{
\begin{split}  
G_{\mathcal{S},1,n}=
\mathrm{Re}\lbs \lb\dfrac{\rho^2 L^2-\omega-2\imag\rho L\sqrt{\omega /2}}{\lb\rho L+\sqrt{\omega /2}\rb^2+\omega/2}\rb^n e^{\imag\omega t} \rbs
\\+\dfrac{1}{\pi}\int_0^\infty\mathrm{Im}\lbs\dfrac{R\lb\rho L-\imag\sqrt{R}\rb^{2n}}{\lb R^2+\omega^2\rb\lb\rho^2 L^2+R\rb^n}\rbs e^{-Rt} \mathrm{d}R.
\end{split}
}
We introduce parameters $\mathcal{K}_{0-3}$ and separate the real and imaginary parts of the complex exponentials and the numerators in each term:
\eq{G_{\mathcal{S},0,n}=
\dfrac{\mathcal{K}_1}{\mathcal{K}_0^n}\sin{\lb\omega t\rb}+\dfrac{\mathcal{K}_2}{\mathcal{K}_0^n}\cos{\lb\omega t\rb}
-\dfrac{1}{\pi}\int_0^\infty \mathcal{K}_3 \dfrac{\omega e^{-Rt}}{R^2+\omega^2} \mathrm{d}R,
}
\eq{G_{\mathcal{S},1,n}=
\dfrac{\mathcal{K}_1}{\mathcal{K}_0^n}\cos{\lb\omega t\rb}-\dfrac{\mathcal{K}_2}{\mathcal{K}_0^n}\sin{\lb\omega t\rb}
+\dfrac{1}{\pi}\int_0^\infty \mathcal{K}_3 \dfrac{R e^{-Rt}}{R^2+\omega^2} \mathrm{d}R,
}
where

\begin{tabular}{cc}
  $\mathcal{K}_0=\lb\rho L+\sqrt{\omega /2}\rb^2+\omega/2,$&
  $\mathcal{K}_1=\mathrm{Re}\lbs\lb \rho^2 L^2-\omega-2\imag\rho L\sqrt{\omega /2} \rb^n\rbs,$\\
  $\mathcal{K}_2=\mathrm{Im}\lbs\lb \rho^2 L^2-\omega-2\imag\rho L\sqrt{\omega /2} \rb^n\rbs,$&
  $\mathcal{K}_3\lb R \rb=\mathrm{Im}\lbs \dfrac{\lb\rho L-\imag\sqrt{R}\rb^{2n}}{\lb\rho^2 L^2+R\rb^n}\rbs.$
\end{tabular}

It is clear that the enclosed residues at $s=\pm \imag\omega$ correspond to the time-periodic terms in $G_{\mathcal{S},m,n}$ and the integrals $\mathrm{I_C}$ and $\mathrm{I_E}$ along either side of the branch cut correspond to the transient terms.

\subsection*{Contour integrals for $G_{\mathcal{F},m,n}$}
On $C$, $s=Re^{\imag \theta}$ as $\theta \to \pi$. Hence, $s=-R$, $\sqrt{s}=\imag\sqrt{R}$ and $\mathrm{d}s=-\mathrm{d}R$:
\eq{\mathrm{I_C} = \int_\infty^0 \dfrac{2 \rho L \lb-R\rb^{1+m} \omega^{1-m} e^{\imag\sqrt{R}\lb h_{\delta}-y_{\delta}\rb}}{\lb R^2+\omega^2\rb \lb \rho L + \imag\sqrt{R}\rb} \lb \dfrac{\rho L-\imag\sqrt{R}}{\rho L+\imag\sqrt{R}} \rb^n e^{-Rt} \lb -\mathrm{d}R \rb.}
On $E$, $s=Re^{\imag \theta}$ as $\theta \to -\pi$. Hence, $s=-R$, $\sqrt{s}=-\imag\sqrt{R}$ and $\mathrm{d}s=-\mathrm{d}R$:
\eq{\mathrm{I_E} = \int_0^\infty \dfrac{2 \rho L  \lb-R\rb^{1+m} \omega^{1-m} e^{-\imag\sqrt{R}\lb h_{\delta}-y_{\delta}\rb}}{\lb R^2+\omega^2\rb \lb \rho L - \imag\sqrt{R}\rb} \lb \dfrac{\rho L+\imag\sqrt{R}}{\rho L-\imag\sqrt{R}} \rb^n e^{-Rt} \lb -\mathrm{d}R \rb,}
\begin{align*}
    \therefore \mathrm{I_C}+\mathrm{I_E}=\int_0^\infty \dfrac{2\rho L\lb-R\rb^{1+m}\omega^{1-m}}{R^2+\omega^2} & \lb\lb \dfrac{\rho L-\imag\sqrt{R}}{\rho L+\imag\sqrt{R}} \rb^n\dfrac{e^{\imag\lb h_{\delta}- y_{\delta}\rb \sqrt{R}}}{\rho L+\imag\sqrt{R}}\right.\\-&
  \left.\lb \dfrac{\rho L+\imag\sqrt{R}}{\rho L-\imag\sqrt{R}} \rb^n\dfrac{e^{-\imag\lb h_{\delta}- y_{\delta}\rb \sqrt{R}}}{\rho L-\imag\sqrt{R}}\rb e^{-Rt}\mathrm{d}R.
\end{align*}
On $D$, $s=R_2 e^{\imag \theta}$. Hence $\sqrt{s}=e^{\imag\theta /2}\sqrt{R_2}$ and $\mathrm{d}s=\imag R_2 e^{\imag \theta}\mathrm{d}\theta$:
\eq{\mathrm{I_D} = \int_\pi^{-\pi} \dfrac{2 \rho L  R_2^{1+m} e^{\lb 1+m \rb \imag\theta} \omega^{1-m}  e^{e^{\imag \theta /2}\sqrt{R_2} \lb h_{\delta}-y_{\delta}\rb}}{\lb  R_2^2 e^{2 \imag \theta}+\omega^2\rb \lb \rho L +  e^{\imag \theta /2}\sqrt{R_2}\rb} \lb \dfrac{\rho L- e^{\imag \theta /2}\sqrt{R_2}}{\rho L+ e^{\imag \theta /2}\sqrt{R_2}}  \rb ^n e^{R_2 e^{\imag\theta}t} \imag R_2 e^{\imag\theta} \mathrm{d}\theta.}

When constructing the solution $G_{\mathcal{F},m,n}$, either $m=0$ or $m=1$. Hence, $\mathrm{I_D}\to 0$ as $R_2\to 0$. We compute the residues at the two poles $s=\pm \imag\omega$:
\begin{align*}
    \mathrm{Res}_{\lb \imag\omega \rb} &=
    {\lb \dfrac{2 \rho L  s^{1+m} \omega^{1-m} e^{\sqrt{s}\lb h_{\delta}-y_{\delta}\rb}}{\lb  s+\imag\omega \rb \lb \rho L + \sqrt{s}\rb} \lb \dfrac{\rho L - \sqrt{s}}{\rho L + \sqrt{s}}\rb ^n e^{st}\rb}_{s\to \imag\omega} \\
    &=\dfrac{ \rho L  \imag^{m} \omega e^{\sqrt{\imag \omega}\lb h_{\delta}-y_{\delta}\rb}}{\rho L + \sqrt{\imag \omega}} \lb \dfrac{\rho L-\sqrt{\imag\omega}}{\rho L+\sqrt{\imag\omega}} \rb ^n e^{\imag \omega t},
\end{align*}
\begin{align*}
    \mathrm{Res}_{\lb -\imag\omega \rb} &= {\lb \dfrac{2 \rho L  s^{1+m} \omega^{1-m}
    e^{\sqrt{s}\lb h_{\delta}-y_{\delta}\rb}}{\lb  s-\imag\omega \rb \lb \rho L + \sqrt{s}\rb} \lb \dfrac{\rho L - \sqrt{s}}{\rho L + \sqrt{s}}\rb ^n e^{st}\rb}_{s\to -\imag\omega} \\
    &=\dfrac{\rho L  \lb -\imag \rb^{m} \omega  e^{\sqrt{-\imag \omega}\lb h_{\delta}-y_{\delta}\rb}}{\rho L + \sqrt{-\imag \omega}} \lb \dfrac{\rho L-\sqrt{-\imag\omega}}{\rho L+\sqrt{-\imag\omega}} \rb ^n e^{-\imag \omega t}.
\end{align*}

The solutions are:
\eq{
  \begin{split}
    G_{\mathcal{F},0,n}=\dfrac{\rho L\omega e^{\lb h_{\delta}-y_{\delta} \rb \sqrt{\imag\omega}}}{\rho L+\sqrt{\imag\omega}}\lb\dfrac{\rho L-\sqrt{\imag\omega}}{\rho L+\sqrt{\imag\omega}}\rb^n e^{\imag\omega t}+\dfrac{\rho L\omega e^{\lb h_{\delta}-y_{\delta} \rb \sqrt{-\imag\omega}}}{\rho L+\sqrt{-\imag\omega}}\lb\dfrac{\rho L-\sqrt{-\imag\omega}}{\rho L+\sqrt{-\imag\omega}}\rb^n e^{-\imag\omega t}&\\
    +\dfrac{1}{2\pi \imag}\int_0^\infty \dfrac{2\rho L R\omega}{R^2+\omega^2}\left(\lb\dfrac{\rho L-\imag\sqrt{R}}{\rho L+\imag\sqrt{R}}\rb^n\dfrac{e^{\imag\lb h_{\delta}-y_{\delta} \rb \sqrt{R}}}{\rho L+\imag\sqrt{R}}\right. \left.-\lb\dfrac{\rho L+\imag\sqrt{R}}{\rho L-\imag\sqrt{R}}\rb^n\dfrac{e^{-\imag\lb h_{\delta}-y_{\delta} \rb \sqrt{R}}}{\rho L-\imag\sqrt{R}}\right)e^{-Rt}\mathrm{d}R&,
\end{split}
\label{contour-output-GF0}
}
\eq{
  \begin{split}
  G_{\mathcal{F},1,n}=\dfrac{\rho L \imag\omega e^{\lb h_{\delta}-y_{\delta} \rb \sqrt{\imag\omega}}}{\rho L+\sqrt{\imag\omega}}\lb\dfrac{\rho L-\sqrt{\imag\omega}}{\rho L+\sqrt{\imag\omega}}\rb^n e^{\imag\omega t}-\dfrac{\rho L \imag\omega e^{\lb h_{\delta}-y_{\delta} \rb \sqrt{-\imag\omega}}}{\rho L+\sqrt{-\imag\omega}}\lb\dfrac{\rho L-\sqrt{-\imag\omega}}{\rho L+\sqrt{-\imag\omega}}\rb^n e^{-\imag\omega t}&\\
  -\dfrac{1}{2\pi \imag}\int_0^\infty \dfrac{2\rho L R^2}{R^2+\omega^2}\left(\lb\dfrac{\rho L-\imag\sqrt{R}}{\rho L+\imag\sqrt{R}}\rb^n\dfrac{e^{\imag\lb h_{\delta}-y_{\delta} \rb \sqrt{R}}}{\rho L+\imag\sqrt{R}}\right.\left.-\lb\dfrac{\rho L+\imag\sqrt{R}}{\rho L-\imag\sqrt{R}}\rb^n\dfrac{e^{-\imag\lb h_{\delta}-y_{\delta} \rb \sqrt{R}}}{\rho L-\imag\sqrt{R}}\right)e^{-Rt}\mathrm{d}R&.
\end{split}
\label{contour-output-GF1}
}
As complex conjugates appear, the solutions are written as:
\eq{
  \begin{split}
  G_{\mathcal{F},0,n}=&
  2 \mathrm{Re}\lbs\dfrac{\rho L\omega e^{\lb h_{\delta}-y_{\delta} \rb \sqrt{\imag\omega}}}{\rho L+\sqrt{\imag\omega}}\lb\dfrac{\rho L-\sqrt{\imag\omega}}{\rho L+\sqrt{\imag\omega}}\rb^n e^{\imag\omega t}\rbs\\
  &+\dfrac{1}{2\pi \imag}\int_0^\infty \imag \dfrac{4\rho L R\omega}{R^2+\omega^2}\mathrm{Im}\lbs\lb\dfrac{\rho L-\imag\sqrt{R}}{\rho L+\imag\sqrt{R}}\rb^n\dfrac{e^{\imag\lb h_{\delta}-y_{\delta} \rb \sqrt{R}}}{\rho L+\imag\sqrt{R}}\rbs e^{-Rt}\mathrm{d}R,
\end{split}
}
\eq{
  \begin{split}
  G_{\mathcal{F},1,n}=&
  2 \mathrm{Re}\lbs\dfrac{\rho L \imag\omega e^{\lb h_{\delta}-y_{\delta} \rb \sqrt{\imag\omega}}}{\rho L+\sqrt{\imag\omega}}\lb\dfrac{\rho L-\sqrt{\imag\omega}}{\rho L+\sqrt{\imag\omega}}\rb^n e^{\imag\omega t}\rbs\\
  &-\dfrac{1}{2\pi \imag}\int_0^\infty \imag \dfrac{4\rho L R^2}{R^2+\omega^2}\mathrm{Im}\lbs\lb\dfrac{\rho L-\imag\sqrt{R}}{\rho L+\imag\sqrt{R}}\rb^n\dfrac{e^{\imag\lb h_{\delta}-y_{\delta} \rb \sqrt{R}}}{\rho L+\imag\sqrt{R}}\rbs e^{-Rt}\mathrm{d}R.
\end{split}
}
We rewrite the denominators of the complex fractions to isolate the overall real and imaginary components:
\eq{
  \begin{split}
  G_{\mathcal{F},0,n}=&
  2 \mathrm{Re}\lbs
  \dfrac{\rho L \omega e^{\lb h_{\delta}-y_{\delta} \rb \sqrt{\omega /2}} \lb\rho^2 L^2 -\omega-2\imag\rho L\sqrt{\omega /2}\rb^n}{\lb\lb\rho L+\sqrt{\omega /2}\rb^2+\omega/2\rb^{n+1}}\lb\rho L+\sqrt{-\imag\omega}\rb e^{\imag\lb\sqrt{\omega/2}\lb h_\delta-y_\delta\rb+\omega t\rb}\rbs\\
  &+\dfrac{1}{\pi}\int_0^\infty \dfrac{2\rho L R\omega}{R^2+\omega^2} \mathrm{Im}\lbs\lb\dfrac{\rho L-\imag\sqrt{R}}{\rho L+\imag\sqrt{R}}\rb^n\dfrac{e^{\imag\lb h_{\delta}-y_{\delta} \rb \sqrt{R}}}{\rho^2 L^2+R} \lb \rho L-\imag\sqrt{R} \rb\rbs e^{-Rt}\mathrm{d}R,
\end{split}
}
\eq{
  \begin{split}
  G_{\mathcal{F},1,n}=&
  2 \mathrm{Re}\lbs
  \dfrac{\rho L \imag \omega e^{\lb h_{\delta}-y_{\delta} \rb \sqrt{\omega /2}} \lb\rho^2 L^2 -\omega-2\imag\rho L\sqrt{\omega /2}\rb^n}{\lb\lb\rho L+\sqrt{\omega /2}\rb^2+\omega/2\rb^{n+1}}\lb\rho L+\sqrt{-\imag\omega}\rb e^{\imag\lb\sqrt{\omega/2}\lb h_\delta-y_\delta\rb+\omega t\rb}\rbs\\
  &-\dfrac{1}{\pi}\int_0^\infty \dfrac{2\rho L R^2}{R^2+\omega^2}\mathrm{Im}\lbs\lb\dfrac{\rho L-\imag\sqrt{R}}{\rho L+\imag\sqrt{R}}\rb^n\dfrac{e^{\imag\lb h_{\delta}-y_{\delta} \rb \sqrt{R}}}{\rho^2 L^2+R} \lb \rho L-\imag\sqrt{R} \rb\rbs e^{-Rt}\mathrm{d}R.
\end{split}
}
We introduce parameters $\mathcal{K}_{4-7}$ and separate the real and imaginary parts of the complex exponentials and the numerators in each term:
\eq{
\begin{split}
  G_{\mathcal{F},0,n}=&\dfrac{2\rho L\omega e^{\lb h_{\delta}-y_{\delta} \rb \sqrt{\omega /2}}}{\mathcal{K}_0^{n+1}} \lb \mathcal{K}_4 \cos{\lb\lb h_{\delta}-y_{\delta} \rb \sqrt{\omega /2}+\omega t\rb} +\mathcal{K}_5 \sin{\lb\lb h_{\delta}-y_{\delta} \rb \sqrt{\omega /2}+\omega t\rb} \rb \\
  &+\dfrac{1}{\pi}\int_0^\infty \dfrac{2\rho L\omega R e^{-Rt}}{\lb R^2 + \omega^2\rb\lb\rho^2 L^2 +R\rb}\lb \mathcal{K}_6\sin{\lb\lb h_{\delta}-y_{\delta} \rb \sqrt{R}\rb}+\mathcal{K}_7\cos{\lb\lb h_{\delta}-y_{\delta} \rb \sqrt{R}\rb}\rb\mathrm{d}R,
\end{split}
}
\eq{
\begin{split}  
  G_{\mathcal{F},1,n}=&\dfrac{2\rho L\omega e^{\lb h_{\delta}-y_{\delta} \rb \sqrt{\omega /2}}}{\mathcal{K}_0^{n+1}} \lb \mathcal{K}_5 \cos{\lb\lb h_{\delta}-y_{\delta} \rb \sqrt{\omega /2}+\omega t\rb} -\mathcal{K}_4 \sin{\lb\lb h_{\delta}-y_{\delta} \rb \sqrt{\omega /2}+\omega t\rb} \rb \\&-\dfrac{1}{\pi}\int_0^\infty \dfrac{2\rho L R^2 e^{-Rt}}{\lb R^2 + \omega^2\rb\lb\rho^2 L^2 +R\rb}\lb \mathcal{K}_6\sin{\lb\lb h_{\delta}-y_{\delta} \rb \sqrt{R}\rb}+\mathcal{K}_7\cos{\lb\lb h_{\delta}-y_{\delta} \rb \sqrt{R}\rb}\rb\mathrm{d}R,
\end{split}
}
where
\eq{\mathcal{K}_4=\lb\rho L+\sqrt{\omega /2}\rb\mathrm{Re}\lbs \lb\rho^2 L^2 -\omega -2\imag \rho L \sqrt{\omega /2}\rb^n \rbs+\sqrt{\omega /2}\mathrm{Im}\lbs \lb\rho^2 L^2 -\omega -2\imag \rho L \sqrt{\omega /2}\rb^n \rbs,}
\eq{\mathcal{K}_5=\sqrt{\omega /2}\mathrm{Re}\lbs \lb\rho^2 L^2 -\omega -2\imag \rho L \sqrt{\omega /2}\rb^n \rbs-\lb\rho L+\sqrt{\omega /2}\rb\mathrm{Im}\lbs \lb\rho^2 L^2 -\omega -2\imag \rho L \sqrt{\omega /2}\rb^n \rbs,}
\eq{\mathcal{K}_6=\rho L \mathrm{Re}\lbs \lb\dfrac{\rho L-\imag\sqrt{R}}{\rho L+\imag\sqrt{R}}\rb^n \rbs+\sqrt{R}\mathrm{Im}\lbs \lb\dfrac{\rho L-\imag\sqrt{R}}{\rho L+\imag\sqrt{R}}\rb^n \rbs,}
\eq{\mathcal{K}_7=\rho L \mathrm{Im}\lbs \lb\dfrac{\rho L-\imag\sqrt{R}}{\rho L+\imag\sqrt{R}}\rb^n \rbs-\sqrt{R}\mathrm{Re}\lbs \lb\dfrac{\rho L-\imag\sqrt{R}}{\rho L+\imag\sqrt{R}}\rb^n \rbs.}

As with the solid-layer solution, the enclosed residues at $s=\pm \imag\omega$ correspond to the time-periodic terms in $G_{\mathcal{F},m,n}$ and the integrals $\mathrm{I_C}$ and $\mathrm{I_E}$ along either side of the branch cut correspond to the transient terms.

%% file: include/B-finite-difference.tex
\section{Numerical procedures}
\label{fd-appendix}
In order to compute the numerical solution used in figure \ref{numerical-checks}, equations \eref{FD-PDE-solid} and \eref{FD-PDE-fluid} are discretised in time with an index $j\geq 0$ and step size $\Delta t$, and in space with an index $0\leq k \leq K$ and step sizes $\Delta y_{\lambda}$ and $\Delta y_{\delta}$ for the solid and fluid, respectively. The interface is characterised by $k=I_s$ and $k=I_f=I_s+1$. For $j \geq 2$, the following approximations are used for \eref{eq1} - \eref{bc4}:

$k=0:$
\eq{
    \begin{array}{ccc}
        \varphi _0^j=\sin (\omega\,j\Delta t), &\qquad\qquad\qquad&
        \bar{\varphi }_0^j=\frac{\varphi _1^j-\varphi _0^j}{\Delta y_{\lambda}}.
    \end{array}
}
$1\leq k <I_s:$
\eq{
    \begin{array}{ccc}
        \frac{\varphi_{k+1}^j-2\varphi_{k}^j+\varphi_{k-1}^j}{2\lb\Delta y_{\lambda}\rb^2}+\frac{\varphi_{k+1}^{j-2}-2\varphi_{k}^{j-2}+\varphi_{k-1}^{j-2}}{2\lb\Delta y_{\lambda}\rb^2}=\frac{\varphi _k^{j-2}-2 \varphi _k^{j-1}+\varphi _k^j}{\lb\Delta t\rb^2},
        &\qquad\qquad&
        \bar{\varphi }_k^j=\frac{\varphi _{k+1}^j-\varphi _{k-1}^j}{2 \Delta y_{\lambda}},
    \end{array}
}
$k=I_s$ and $k=I_f:$
\eq{
    \begin{array}{ccc}
        \frac{\bar{\varphi }_{I_s}^j-\bar{\varphi }_{I_s-1}^j}{\Delta y_{\lambda}}=\frac{\varphi _{I_s}^{j-2}-2 \varphi _{I_s}^{j-1}+\varphi_{I_s}^j}{\lb\Delta t\rb^2},
        &\qquad\qquad\qquad&
        \bar{\varphi }_{I_s}^j=\frac{\varphi _{I_s}^j-\varphi _{I_s-1}^j}{\Delta y_{\lambda}},
    \end{array}
}
\eq{
    \begin{array}{ccc}
        \frac{\bar{u}_{I_f+1}^j-\bar{u}_{I_f}^j}{\Delta y_{\delta}}=\frac{u_{I_f}^{j-2}-4 u_{I_f}^{j-1}+3 u_{I_f}^j}{2\Delta t},
        &\qquad\qquad\qquad&
        \bar{u}_{I_f}^j=\frac{u_{I_f+1}^j-u_{I_f}^j}{\Delta y_{\delta}},
    \end{array}
}
\eq{
    \begin{array}{ccc}
        u_{I_f}^j=\frac{\varphi _{I_s}^{j-2}-4 \varphi _{I_s}^{j-1}+3\varphi _{I_s}^j}{2\Delta t},
        &\qquad\qquad\qquad&
        \bar{u}_{I_f}^j=\rho L \bar{\varphi }_{I_s}^j.
    \end{array}
}
$I_f<k<K:$
\eq{
    \begin{array}{ccc}
        \frac{\bar{u}_{k+1}^j-\bar{u}_{k-1}^j}{2 \Delta y_{\delta}}=\frac{u_k^{j-2}-4 u_k^{j-1}+3u_k^j}{2\Delta t},
        &\qquad\qquad\qquad&
        \bar{u}_k^j=\frac{u_{k+1}^j-u_{k-1}^j}{2 \Delta y_{\delta}}.
    \end{array}
}
$k=K:$
\eq{
    \begin{array}{ccc}
        u_K^j=0,
        &\qquad\qquad\qquad&
        \bar{u}_K^j=\frac{u_K^j-u_{K-1}^j}{\Delta y_{\delta}}.
    \end{array}
}